\documentclass[twocolumn,10pt]{article}

\usepackage{amsmath}
\usepackage{amssymb}
\usepackage{graphicx}
\usepackage{algorithm}
\usepackage{algpseudocode}
\usepackage{bm}
\usepackage[hidelinks]{hyperref}
\usepackage{xcolor}
\usepackage{siunitx}
\usepackage{booktabs}
\usepackage{multirow}
\usepackage[margin=1in]{geometry}
\usepackage{authblk}
\usepackage{abstract}
\usepackage{float}
\usepackage{dblfloatfix}
\usepackage[numbers,sort&compress]{natbib}

\newcommand{\vb}{\mathbf{b}}
\newcommand{\vn}{\mathbf{n}}
\newcommand{\vr}{\mathbf{r}}
\newcommand{\vt}{\bm{\xi}}

\begin{document}

\title{Savi-Bhransha: Graph-Theoretic Dislocation-Loop Characterization in Crystals}

\author[1,2]{Utkarsh Bhardwaj\thanks{utkarsh@barc.gov.in}}

\affil[1]{Computational Analysis Division, Bhabha Atomic Research Centre, Visakhapatnam, Andhra Pradesh, India 530012}
\affil[2]{Homi Bhabha National Institute, Anushaktinagar, Mumbai, Maharashtra, India 400094}

\date{}

\twocolumn[
\maketitle
\begin{onecolabstract}
Dislocation loops govern the properties of crystalline materials, but extracting their detailed characteristics from atomistic simulations is difficult when loops are fragmented or embedded in compact defect debris. We present Savi-Bhransha, a graph-theoretic method that reconstructs interstitial and vacancy loops directly from local defect-displacement motifs, without constructing a global interface mesh. The method identifies Burgers-vector family, habit plane, loop size, segment-wise edge/screw character, and boundary and bulk defect populations for BCC, FCC, and HCP crystals. We apply it to single-cascade simulations over a range of energies in BCC W and HCP Zr, and to successive collision cascades in BCC W and FCC FeNiCr. We benchmark the method against the Dislocation Extraction Algorithm (DXA). Total dislocation lengths remain strongly correlated between the two methods, while Savi-Bhransha returns more stable loop-level objects in complex environments where DXA returns fragmented, overlapping open segments. Savi-Bhransha also better resolves mixed-morphology defects and dislocations near other defects, including vacancy clusters. Median runtime speedups are $6.34\times$ for BCC W and $8.86\times$ for HCP Zr, with peak-memory reductions up to $7.77\times$. In successive W cascades, the resolved boundary-defect concentration brackets transient-grating-spectroscopy measurements and the predicted Burgers-vector fraction agrees with room-temperature TEM. In FCC FeNiCr, the method resolves Heidenreich--Shockley dissociation, with a Shockley-pair signature in about 91\% of surviving $\langle110\rangle$-family interstitial clusters. Savi-Bhransha therefore enables efficient, topology-resolved analysis of large radiation-damage simulations and direct comparison with experimentally accessible observables.
\end{onecolabstract}
]


\section{Introduction}
\label{sec:intro}

Defects and more specifically dislocation loops govern the mechanical, thermal, and transport response of crystalline materials. In metals, their size, Burgers-vector population, vacancy-versus-interstitial character, and spatial arrangement control swelling, hardening, defect mobility, and transport degradation~\cite{bacon2000,nordlund2018}. Non-dislocation defect structures such as point defects and their aggregates, compact three-dimensional clusters, and C15-like motifs can pin or immobilize otherwise mobile loops and alter subsequent microstructural evolution~\cite{sand2013,setyawan2015}. Atomistic simulations must therefore be reduced not only to defect densities but to object-level morphology---including internal loop structure, the surrounding defect environment, and segment-wise dislocation character---to enable meaningful comparison with mesoscale models and experimental observables.

Molecular dynamics (MD) simulations resolve defect production, migration, and interactions, and carry rich information on the underlying mechanisms required for multi-scale modelling of property changes under extreme conditions such as the high-dose irradiation environments of interest in fusion, aerospace, and related fields. In practice, however, the raw atomic trajectories are uninformative on their own: extracting that information requires analysis algorithms capable of reducing millions of atomic coordinates to discrete defect objects with well-defined topology and morphology. Continuum descriptors such as the Nye tensor capture local defect-density fields but do not directly identify discrete loop objects. The Dislocation Extraction Algorithm (DXA)~\cite{stukowski2010dxa,stukowski2012dxa} has become a standard tool for recovering continuous dislocation lines from atomistic configurations by constructing Burgers circuits on an interface mesh~\cite{fikar2018vacancy,derlet2020}. The algorithm scales with the total number of atoms, which becomes costly for large cascade simulation boxes with sparse defect populations. Moreover, cascade debris is not always a clean dislocation-network environment: loops may be embedded in compact clusters, vacancy-rich zones, or C15-like structures. In such mixed-cluster contexts the defect morphology of composite clusters is harder to recover, neighbouring defects in the vicinity can cause DXA to return a single physical loop as several open segments, and smaller constituents of the cluster---both small dislocation and non-dislocation components that may be pinning the loop---are not always reflected in its native output.

These limitations matter because the missing information is often the physically relevant information. A loop entangled with compact debris may have different mobility from an isolated loop, even if its Burgers vector is the same; small vacancy loops can be important even when they are difficult to recover from a dislocation-line mesh~\cite{fikar2018vacancy}; and local dumbbell arrangements near loop interfaces affect loop stability and evolution~\cite{bhardwaj2022loop}. The distinction between defects at a loop perimeter and defects in the loop interior is also essential for thermal-transport comparisons: transient grating spectroscopy (TGS) measurements of thermal diffusivity are sensitive primarily to the boundary defect population that scatters phonons, and prior cascade studies report improved agreement with experiment once boundary and bulk populations are separated~\cite{dennett2019tgs,reza2020tgsthermal,mason2021thermal,bhardwaj2024scc}.

To resolve defect morphology in finer detail, we previously developed SAVi (Samuh AnuVikar), a graph-theoretic scheme in which dumbbell and crowdion configurations form the nodes and crystallographic relationships between them form the edges~\cite{bhardwaj2021graph}. Building on that foundation, the present work introduces Savi-Bhransha (\textit{Bhransha}: Sanskrit for ``dislocation''), which extends SAVi to recover full loop topology, Burgers vector, and habit plane directly from the defect displacement field and the crystallographic relationships between its components. Encoding these constraints locally in the adjacency graph removes the need for a global interface mesh, yielding a complete dislocation analysis at higher computational efficiency.

Savi-Bhransha operates in the inverse direction to conventional dislocation construction~\cite{mura1987,barnett1985,cai2006nonsingular,hirel2015atomsk}, recovering the loop topology from the high-displacement core configuration. The atomic configuration already contains the displacement field produced by the interatomic potential; the algorithm identifies local dumbbell, displaced atom--vacancy, and vacancy-centered motifs, assigns them to crystallographic direction families, and merges them into extended line primitives. A sparse adjacency graph over these primitives---built without any global atomic mesh and applicable uniformly to BCC, FCC, and HCP lattices---then groups primitives that share similar distance and angle relationships, and identifies the interstitial and vacancy dislocation loops whose displacement-field signatures match those of a dislocation. The same construction also resolves C15-like structures and mixed morphologies within a single pipeline. Geometric coarse-graining of each component yields quantitative loop descriptors: Burgers-vector family, habit plane with confidence metrics, loop perimeter and equivalent diameter, segment-wise edge/screw character, and an explicit boundary-versus-bulk defect separation.

We demonstrate Savi-Bhransha on radiation-damage cascade microstructures, a setting that combines dislocation loops with compact debris, vacancy-rich zones, and C15-like motifs and is representative of dislocation analysis in complex defect environments more generally. The method is applied to single-PKA BCC~W and HCP~Zr low to high energy sweeps, successive 50\,keV cascades in BCC~W up to 0.2\,dpa, and a 20\,keV successive-cascade FCC~FeNiCr trajectory. In BCC~W at 0.1\,dpa, the resolved boundary-defect concentration ($1.8$--$2.7\times10^{-3}$ across two interatomic potentials) brackets the TGS-measured value of $2.25\times10^{-3}$, supporting a boundary-driven interpretation of the measured thermal-diffusivity drop, and the SNAP $\frac{1}{2}\langle111\rangle$ fraction at 0.20\,dpa matches the room-temperature TEM value of Yi et~al.\ within experimental uncertainty. In FCC~FeNiCr, the line-primitive graph resolves Heidenreich--Shockley dissociation of $\frac{1}{2}\langle110\rangle$ loops into $\frac{1}{6}\langle112\rangle$ Shockley-partial pairs on $\{111\}$ at the population level. Benchmarked against DXA on the BCC~W and HCP~Zr datasets, the two methods agree closely on total dislocation length and on loop populations in the majority of cases; systematic differences emerge in complex defect environments, where DXA can fragment a single physical loop into multiple segments---an effect that requires post-processing to recover loop-level statistics such as loop number density---and where the explicit boundary--bulk separation and the defect-environment detail accessible to Savi-Bhransha are not part of the DXA native output. Across the benchmark, Savi-Bhransha reduces runtime and peak memory by factors approaching an order of magnitude. The remainder of the paper develops the algorithm (Section~\ref{sec:methods}) and presents the DXA benchmark and application studies in turn.

\section{Methodology}
\label{sec:methods}

Forward dislocation construction in continuum elasticity expresses the displacement field of a loop as a Mura--Willis surface integral, evaluated for closed polygonal loops via Barnett's straight-segment decomposition with the Burgers vector and habit plane as inputs~\cite{mura1987,barnett1985}; non-singular continuum theories regularise the resulting core singularity by spreading the Burgers vector over a finite radius~\cite{cai2006nonsingular}, and the same forward construction underlies modern dislocation-generation tools~\cite{hirel2015atomsk}. DXA, by contrast, recovers an existing dislocation network from an atomistic configuration by constructing Burgers circuits on an interface mesh between defective and crystalline regions~\cite{stukowski2010dxa,stukowski2012dxa}; its theoretical concept is the Burgers-circuit definition of the topological charge rather than the displacement-field construction.

\begin{figure*}[!t]
\centering
\includegraphics[width=0.95\textwidth]{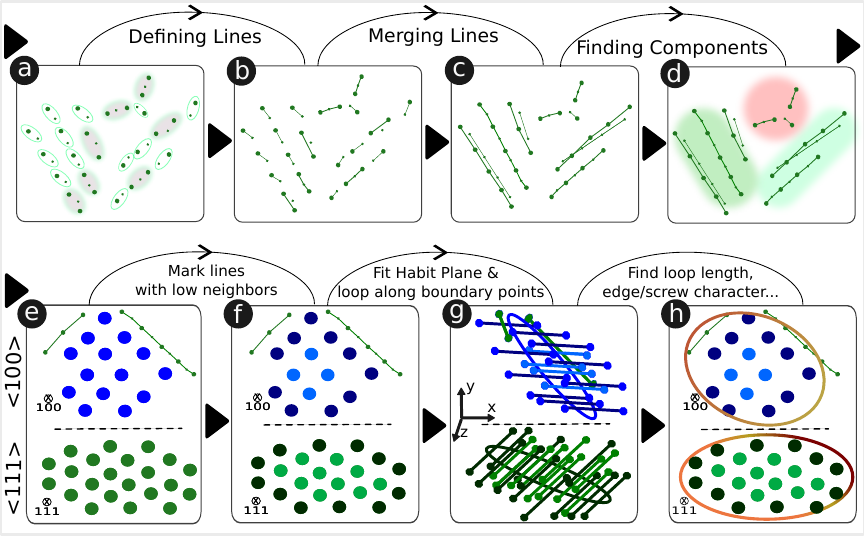}
\caption{Schematic overview of the Savi-Bhransha pipeline. \textit{Top row}: graph construction over line primitives within a single defect cluster. (a)~Defect motifs identified from lattice-site occupancy---isolated dumbbell triads, displaced-atom/vacancy pairs, and crowdions---each contributing a line primitive (Section~\ref{sec:defect_id}). (b)~Line primitives are created by joining the triads and pairs. (c)~Collinear lines are joined to form unified line primitives representing extended crowdions (Section~\ref{sec:merging}). (d)~Connected-component analysis on the type-stratified adjacency graph: sufficiently large parallel components (green shading) correspond to displacement fields associated with dislocation loops, and ring components (red shading) correspond to C15-like structures (Section~\ref{sec:graph_construct}). \textit{Bottom row}: per-loop dislocation characterisation, illustrated on two example loops in BCC---an $\langle100\rangle$ loop (top, blue) and a $\langle111\rangle$ loop (bottom, green). (e)~Parallel components are initialised with their constituent merged line primitives. (f)~Each line's neighbour count classifies it as boundary or bulk (darker shading: boundary; Section~\ref{sec:boundary}); the boundary-defect count is retained for subsequent analysis and comparison with experiment. (g)~A weighted principal-component fit to the line centroids gives the habit normal, and the boundary primitives define the loop contour (Section~\ref{sec:habit_plane}). (h)~Further properties---the Burgers vector (Section~\ref{sec:burgers_mag}), loop perimeter, equivalent diameter, and segment-wise and total edge/screw character (Section~\ref{sec:character})---are then calculated.}
\label{fig:methods_overview}
\end{figure*}

Savi-Bhransha addresses the inverse of the forward-construction problem: given an atomistic configuration whose displacement field has already been produced by the interatomic potential, it recovers the underlying loop topology, Burgers vector, and habit plane directly from the high-displacement core motifs---dumbbells, displaced atom--vacancy pairs, and vacancy-centred triads. The construction is purely geometric and crystallographic, so it requires neither elastic constants nor a Green's function, and the dependence on lattice properties enters only through the symmetry families that constrain the adjacency graph. Figure~\ref{fig:methods_overview} summarises the pipeline schematically; the following subsections develop each stage in order.

\subsection{Defect Identification and Line Primitives}
\label{sec:defect_id}

Given the atomic coordinates of a simulation snapshot, the first step is to
identify the defect content and assign each atom to its nearest ideal lattice
site. Each atom is mapped to its nearest ideal lattice site via modular arithmetic:
given axis periods $\mathbf{p}$ (lattice parameters) and fractional origin $\mathbf{o}$ which is similar to the offset or shift used in building the simulation box,
\begin{equation}
  \tilde{r}_i = (r_i - o_i p_i) \bmod p_i, \qquad
  k^* = \underset{k}{\arg\min}\; d_{\mathrm{PBC}}(\tilde{\mathbf{r}}, \mathbf{s}_k)
\label{eq:modred}
\end{equation}
where $\{\mathbf{s}_k\}$ are the $M$ sub-lattice offsets within one unit cell
($M{=}2$ BCC, $M{=}4$ FCC/HCP) and $d_{\mathrm{PBC}}$ is the minimum-image
distance. The fractional anchor is $a_i = \lfloor(r_i - o_i p_i)/p_i\rfloor
+ s_{k^*,i}/p_i + o_i$.
The resulting fractional anchors are PBC-folded into
$[\mathbf{o},\,\mathbf{o}+N_{\mathrm{cell}})$ and sorted lexicographically.

Lattice-site occupancy is then recovered by co-traversing the sorted anchor list against an on-the-fly enumeration of the ideal sites in the same order, so the full reference lattice is never held in memory. A site absent from the atom list is a vacancy; a site claimed by more than one atom is an interstitial. This recovers the same lattice-site occupancy as the Wigner-Seitz construction~\cite{nordlund1998} without an explicit reference frame or spatial index; total cost is $O(N \log N)$~\cite{anuvikar}. For each over-occupied site we identify the triad: the vacancy
at the lattice position and two displaced atoms forming the dumbbell. The axis
$\mathbf{d}$ connecting the two interstitial atoms defines a line primitive
$\mathcal{L}(\mathbf{d}) = \{\vr_{\text{vac}} + \lambda \mathbf{d} \mid \lambda \in \mathbb{R}\}$.
In addition, atoms whose displacement from their nearest lattice site exceeds a threshold (typically $0.3$--$0.4\times$ the nearest-neighbor spacing~\cite{bukkuru2017sia}) are paired with that lattice site and treated as displaced-atom--vacancy lines. These threshold-based pairs carry zero net defect content, but they capture the diffuse displacement field around extended defects and help in identifying morphological details in later steps. They typically appear in one of three configurations: coincident with a triad-bearing dumbbell line forming a crowdion, coincident with a true vacancy line in vacancy loops, or as non-aligned interface disturbances surrounding a well-defined dislocation or other defect structure.

\subsection{Crystallographic Orientation Assignment}
\label{sec:orient}

Each dumbbell axis $\mathbf{d}$ is assigned to a low-index crystallographic direction family. This orientation governs both graph construction (Section~\ref{sec:graph}) and, once loops are identified, determines the Burgers vector. The identification rests on the Mura--Willis representation of a closed loop of Burgers vector $\mathbf{b}$ bounding a habit surface $S$,
\begin{equation}
\mathbf{u}(\mathbf{r}) = -\frac{\mathbf{b}\,\Omega(\mathbf{r})}{4\pi} + \mathbf{u}_{\mathrm{LI}}(\mathbf{r}),
\label{eq:mura_willis}
\end{equation}
with $\Omega(\mathbf{r})$ the solid angle subtended by $S$ at $\mathbf{r}$ and $\mathbf{u}_{\mathrm{LI}}(\mathbf{r})$ the smooth line-integral remainder governed by the elastic Green's tensor~\cite{mura1987,barnett1985}. The first term is everywhere parallel to $\mathbf{b}$ and carries the entire $\mathbf{b}$-jump across $S$; $\mathbf{u}_{\mathrm{LI}}$ contributes the smaller Poisson-type bulging. The inserted material at the core therefore aligns with $\mathbf{b}$ and is observed atomistically as a dumbbell or crowdion triad parallel to $\mathbf{b}$, so the per-primitive family assignment reads the local Burgers direction directly.

The relevant crystallographic direction families depend on crystal structure:

\textit{BCC:} $\langle 111 \rangle$ (close-packed, produces highly mobile loops when clustered), $\langle 100 \rangle$ (non-close-packed, sessile loops), and $\langle 110 \rangle$ (appears at interfaces between domains or in C15 ring structures); the higher-index $\langle 112 \rangle$ and $\langle 221 \rangle$ families are also optionally retained in the candidate set so that the less common dumbbell orientations encountered in dense cascade debris can still be assigned rather than being forced into a nearby low-index family.

\textit{FCC:} $\langle 110 \rangle$, $\langle 112 \rangle$, $\langle 221 \rangle$, $\langle 100 \rangle$, and $\langle 111 \rangle$ families, enabling separation of perfect and partial/faulted loop populations within the same crystallographic framework.

\textit{HCP:} Three families expressed in Miller-Bravais notation $[UVTW]$:
\begin{itemize}
\item $\langle a \rangle$-type: $\frac{a}{3}[11\bar{2}0]$ and symmetry-equivalent directions;
\item $\langle c \rangle$-type: $[0001]$ along the $c$-axis;
\item $\langle c+a \rangle$-type: $\frac{1}{3}[11\bar{2}3]$ (Type II) and $\frac{1}{3}[10\bar{1}1]$ (Type I).
\end{itemize}
Each dumbbell is assigned to a family using angular matching against pre-computed standard directions, with an integer family code enabling efficient computation in vectorized operations. The families can be added to or removed from the candidate set as needed.

\subsection{Loop Identification: Line Merging and Graph Analysis}
\label{sec:graph}

With orientation families assigned, loop identification proceeds in two stages: collinear defect lines are first merged into extended line primitives, and a graph over these primitives then resolves loops and other morphological units. Both stages operate within a single defect cluster at a time; clusters are obtained beforehand by grouping point defects whose pairwise separation lies within the second-nearest-neighbor (2NN) distance, which keeps every subsequent pairwise computation local.

The two stages share a common pair geometry, parameterized by an inter-line distance and an inter-axis angle. The natural distance measure depends on the relative orientation of the two lines. When axes $\mathbf{d}_i, \mathbf{d}_j$ are nominally parallel (the merging step and the parallel-edge criterion below), we use the perpendicular distance between the axes treated as infinite lines,
\begin{equation}
d_{\perp,ij} = \frac{|(\vr_j - \vr_i) \cdot (\mathbf{d}_i \times \mathbf{d}_j)|}{|\mathbf{d}_i \times \mathbf{d}_j|},
\label{eq:line_dist}
\end{equation}
which reduces to the standard point-to-line distance in the strictly parallel limit. To suppress thermal scatter and local deviations from the ideal orientation, both axes are first projected onto the assigned family direction $\hat{\mathbf{d}}$ before $d_{\perp,ij}$ is evaluated. For non-parallel pairs (the C15 ring criterion), the perpendicular distance between infinite axes is poorly conditioned, and we instead use the underlying point distance between the two primitives---lattice-site centers for SIA-bearing dumbbell lines, vacancy positions for vacancy-centered lines---denoted $\|\vr_i - \vr_j\|$. The relative orientation in both cases is the unsigned angular difference $\theta_{ij} = \arccos(|\hat{\mathbf{d}}_i \cdot \hat{\mathbf{d}}_j|)$.

\subsubsection{Collinear Merging into Line Primitives}
\label{sec:merging}

Individual dumbbells and threshold-based displaced atom--vacancy pairs that share a common axis are merged into unified line segments. Two lines $i$ and $j$ belonging to the same orientation family are merged when
\begin{equation}
d_{\perp,ij} < \delta_{\text{merge}}, \quad \theta_{ij} < \theta_{\text{col}},
\label{eq:merge_criteria}
\end{equation}
where $\delta_{\text{merge}}$ is a tight collinearity threshold---deliberately smaller than the inter-line spacing $\delta_{\perp}$ used later for graph edges---and $\theta_{\text{col}}$ enforces near-parallel alignment. Because of the discrete lattice and thermal vibrations, perfect collinearity for actual line segments is rare but it is possible if the line segments are first assigned a crystallographic family direction and projected onto that. The use of projected line equations and use of tolerances for real line segments ensure that physically collinear defects are represented as single entities.

Merging confers several advantages. First, it absorbs collinear threshold-based lines into the dumbbell triads they accompany---in effect reconstructing crowdion segments---and reduces the primitive count, improving efficiency in subsequent graph analysis. Second, the merged segments yield physically meaningful morphological parameters: segment length varies systematically with position in the loop---shorter at the boundary, longer in the core, reflecting the decay of the displacement field~\cite{derlet2020}---while the number of constituent defects per line constrains the Burgers-vector magnitude. Third, the line centroids, being averages over constituent atoms, provide robust coordinates for habit-plane fitting (Section~\ref{sec:habit_plane}). Finally, the merged lines expose internal morphology: deviations of the fitted axis from the assigned crystallographic direction, offsets between the lattice-site line and the SIA line, and atomic scatter about the axis all serve as signatures of split configurations and local displacement-field structure.

\subsubsection{Graph Construction and Component Analysis}
\label{sec:graph_construct}

On the merged line primitives we construct a graph $G = (V, E)$ in which vertices $V$ are the primitives and edges $E$ are typed by the geometric relationship between their endpoints. Each edge type encodes a different physical relationship between two primitives, and the framework leaves room for additional types as the family of motifs of interest grows; we currently instantiate two. An edge of \textit{parallel} type, appropriate for the loop-forming case where neighbouring lines share a crystallographic direction, is created when
\begin{equation}
d_{\perp,ij} < \delta_{\perp}, \quad \theta_{ij} < \theta_{\text{par}}, \quad \text{family}_i = \text{family}_j,
\label{eq:parallel_edge}
\end{equation}
with $\delta_{\perp}$ scaling with the nearest-neighbor distance (typically $1.5$--$2.0\times$ NN spacing) and $\theta_{\text{par}}$ can be very small as we use crystallographically aligned and projected line equations. An edge of \textit{ring} type, characteristic of C15 configurations whose constituent lines meet at large mutual angles, is created when the underlying point distance between the two primitives is small and the axes are far from parallel:
\begin{equation}
\|\vr_i - \vr_j\| < a\sqrt{2}, \quad \theta_{ij} > 60^\circ.
\label{eq:ring_edge}
\end{equation}

Connected-component analysis is then applied per edge type: nodes joined by parallel edges form a parallel component, and nodes joined by ring edges form a ring component. A node that participates in edges of more than one type sits at the interface between morphological units, and the set of such interface nodes identifies composite or pinned arrangements---a loop entangled with a C15 cluster, for instance, or two loops of different direction families sharing a junction.

Parallel components are identified with dislocation loops, and ring components with C15-like structures or their bases. The identification of a parallel component with a dislocation loop rests on the same Mura--Willis decomposition introduced in Section~\ref{sec:orient}: along the perimeter of a closed loop of Burgers vector $\vb$, the topological displacement contribution is a uniform jump of $\vb$ across the habit plane, and the inserted material that realises this jump appears at the atomistic level as a chain of dumbbell or crowdion triads all aligned with $\vb$. A connected component of mutually parallel line primitives in the graph is therefore the atomistic image of such a perimeter contour, and its Burgers-vector family is set by the common direction-family of its constituents---taken as the majority-vote consensus of their per-primitive labels (Section~\ref{sec:orient}), which absorbs the few peripheral primitives whose family code has been shifted by the Poisson-bulging contribution of the line-integral displacement near the loop perimeter. A parallel cluster of $\langle 111 \rangle$-oriented lines in BCC, for example, thus constitutes a $\frac{a}{2}\langle 111 \rangle$ dislocation loop.

\subsection{Habit Plane Determination}
\label{sec:habit_plane}

The habit plane of a dislocation loop is obtained by a weighted principal-component fit to its constituent line centroids $\{\vr_i\}$, with weights $w_i$ proportional to the constituent-defect count of each line. The weighted covariance
\begin{equation}
C = \frac{\sum_i w_i (\vr_i - \bar{\vr})(\vr_i - \bar{\vr})^\top}{\sum_i w_i}
\label{eq:covariance}
\end{equation}
is diagonalized to give eigenvalues $\lambda_1 \geq \lambda_2 \geq \lambda_3$ and corresponding eigenvectors $\mathbf{e}_\alpha$; the habit normal $\vn = \mathbf{e}_3$ minimizes the weighted sum of squared point-to-plane distances. Two scalar diagnostics accompany the fit: the in-plane scatter
\begin{equation}
\sigma_{\text{RMS}} = \sqrt{\frac{\sum_i w_i [(\vr_i - \bar{\vr}) \cdot \vn]^2}{\sum_i w_i}},
\label{eq:rms}
\end{equation}
and the planarity ratio $R_{\text{planar}} = 1 - \lambda_3/\lambda_1$, which approaches unity for a well-defined planar loop and falls toward zero for three-dimensional or disordered components.

The fitted normal is then matched to a low-index plane family. For each candidate $(hkl)$ the Cartesian normal $\mathbf{n}_{hkl} = B(h,k,l)^\top$ is computed from the reciprocal basis and the deviation $\theta_{\text{dev}} = \arccos(|\vn \cdot \mathbf{n}_{hkl}|)$ is taken as the match metric. Candidate families are $\{110\}, \{111\}, \{100\}$ for BCC and FCC, and $(0001)$ basal, $\{10\bar{1}0\}$ prismatic, $\{10\bar{1}1\}$ pyramidal-I, and $\{11\bar{2}2\}$ pyramidal-II for HCP. Each assignment carries a confidence label derived from the deviation: high ($<5^\circ$), medium ($5^\circ$--$15^\circ$), low ($15^\circ$--$25^\circ$), very low ($>25^\circ$).

\subsection{Burgers Type and Magnitude Inference}
\label{sec:burgers_mag}

With the Burgers family of each loop component already fixed by the connected-component analysis (Section~\ref{sec:graph_construct}), we now refine to a specific Burgers state from the admissible crystallographic set of the parent lattice and report the corresponding magnitude $|\vb|$. Material dependence enters only through lattice geometry and the allowed Burgers set: for BCC and HCP we use the conventional canonical set for the structure, while for FCC the admissible set additionally includes fault-related partial dislocation loops and stacking faults~\cite{jublotleclerc2016,lu2017ris}.

Within a single direction family the specific Burgers state is disambiguated using the resolved habit plane, the discrete defect content carried by the merged line primitives, and, where the local signal is sufficiently clear, the displaced-atom geometry associated with the component. In FCC, for example, both $\frac{1}{2}\langle110\rangle$ and $\frac{1}{6}\langle110\rangle$ states belong to the $\langle110\rangle$ family; their crystallographic distinction is reflected in the resolved loop habit, with perfect loops associated with glide on $\{111\}$ and stair-rod character associated with $\{100\}$ junction geometry.

When the extra atom count per constituent line exceeds unity, it signifies a translation step of more than shortest lattice translation in that direction. The Burgers magnitude follows from the assigned crystallographic type as
\begin{equation}
|\vb| = f\, a_0 \, \|\mathbf{d}\|,
\end{equation}
where $f$ is the fractional coefficient and $\mathbf{d}$ is the primitive direction vector of the corresponding family.

\subsection{Boundary-Bulk Discrimination}
\label{sec:boundary}

Separating defects at the loop perimeter from those in the interior is essential for comparison with thermal-transport measurements, where phonon scattering depends predominantly on the boundary population~\cite{reza2020tgsthermal,bhardwaj2024scc}. For each line $i$ in a parallel component we count its same-family neighbors within the graph-edge distance,
\begin{equation}
n_i = \sum_{j \neq i} \mathbb{1}[d_{\perp,ij} < \delta_{\perp}]\,\mathbb{1}[\text{family}_i = \text{family}_j],
\label{eq:neighbor_count}
\end{equation}
where $\mathbb{1}[\cdot]$ is the indicator function (unity when its condition holds, zero otherwise), so $n_i$ counts the line primitives that are both within the graph-edge distance $\delta_{\perp}$ of line $i$ and share its direction family. We then classify $i$ as \textit{boundary} when $n_i < n_{\text{surf}}$. The threshold reflects the in-plane coordination of a fully embedded line on the relevant habit:
\begin{equation}
n_{\text{surf}} = \begin{cases}
5 & \text{BCC/FCC } \langle 111 \rangle, \text{ HCP } \langle c \rangle/\langle c{+}a \rangle, \\
4 & \text{BCC/FCC } \langle 100 \rangle, \text{ HCP } \langle a \rangle,
\end{cases}
\label{eq:surf_thresh}
\end{equation}
so that close-packed and basal-$\langle a\rangle$ loops carry their respective characteristic interior coordinations. The resulting boundary count $N_{\text{boundary}}$ and bulk count $N_{\text{bulk}} = N_{\text{total}} - N_{\text{boundary}}$ are object-level descriptors that can be compared directly with the boundary-sensitive observables of TGS and related experiments.

\subsection{Dislocation Character Analysis}
\label{sec:character}

With the Burgers vector $\vb$ and habit normal $\vn$ in hand, the edge/screw character is resolved along the loop perimeter. The boundary defect coordinates are first projected onto the habit plane,
\begin{equation}
\vr'_i = \vr_i - [(\vr_i - \bar{\vr}) \cdot \vn]\vn,
\label{eq:project}
\end{equation}
and expressed in an in-plane orthonormal basis $(\mathbf{u}_1, \mathbf{u}_2)$ as $(x_i, y_i) = (\vr'_i \cdot \mathbf{u}_1, \vr'_i \cdot \mathbf{u}_2)$. An $\alpha$-shape (concave hull) is then fitted to these 2D points, with $\alpha$ chosen tight enough to track the interatomic spacing of cascade defects, yielding an ordered perimeter sequence $\{\vr'_k\}$. The perimeter and enclosed area follow as
\begin{equation}
P = \sum_k |\vr'_{k+1} - \vr'_k|, \quad A = \frac{1}{2}\left|\sum_k (x_k y_{k+1} - x_{k+1} y_k)\right|,
\label{eq:perimeter_area}
\end{equation}
and the equivalent circular diameter $d_{\text{circ}} = 2\sqrt{A/\pi}$ summarizes the loop size.

For each perimeter segment, the local tangent $\vt_k = (\vr'_{k+1} - \vr'_k)/|\vr'_{k+1} - \vr'_k|$ defines the unsigned angle $\phi_k \in [0, 90^\circ]$ with the Burgers vector via $\cos\phi_k = |\vt_k \cdot \hat{\vb}|$ (the absolute value renders the decomposition invariant to traversal direction). The classical dislocation character decomposition then gives
\begin{equation}
f_{\text{screw},k} = \cos^2\phi_k, \quad f_{\text{edge},k} = \sin^2\phi_k,
\label{eq:fractions}
\end{equation}
with $f_{\text{screw}} + f_{\text{edge}} = 1$. Segments are labeled screw ($f_{\text{screw}} > 0.8$), edge ($f_{\text{edge}} > 0.8$), or mixed, and a length-weighted mean
\begin{equation}
\bar{f}_{\text{edge}} = \frac{\sum_k L_k f_{\text{edge},k}}{\sum_k L_k},\qquad L_k = |\vr'_{k+1} - \vr'_k|,
\label{eq:avg_char}
\end{equation}
characterizes the overall loop.

\subsection{Partial dislocations in FCC}
\label{sec:shockley}

In FCC metals with low stacking-fault energy, a perfect $\frac{1}{2}\langle110\rangle\{111\}$ dislocation spontaneously dissociates into two Shockley partials of Burgers vector $\frac{1}{6}\langle112\rangle$ bounded by a ribbon of intrinsic stacking fault~\cite{hirth1982}. The object-level loop descriptors do not by themselves separate the dissociated state from the undissociated one: both carry the same net $\frac{1}{2}\langle110\rangle$ Burgers sum, and the dominant line-direction family inferred from the triad primitives can remain $\langle110\rangle$ when the two partials sit close together. An optional analysis stage resolves the partial structure where it is present.

The two partial-dislocation cores that bound the stacking fault sit on two distinct $\{111\}$ layers, populated by dumbbell or crowdion triads with surviving displaced atoms; the stacking-fault ribbon between them carries no inserted material and is signalled instead by non-surviving displaced-atom and vacancy pairs. Operationally, the dumbbells of each $\langle110\rangle$-family SIA cluster are partitioned into bands across the candidate $\{111\}$ habit normal, and the partial structure is read from the populations of those bands: a single populated band keeps the perfect $\frac{1}{2}\langle110\rangle$ label; two well-populated outer bands (with an optional sparse middle band from transient SF-region atoms) are reported as a Shockley pair emitting two $\frac{1}{6}\langle112\rangle$ partials on $\{111\}$; and three distinct $\{111\}$ planes with largely disjoint band populations identify a stair-rod junction emitting two Shockley pairs together with a $\frac{1}{6}\langle110\rangle$ stair-rod along the intersection line.

A residual subset of clusters has all vacancy lattice sites coplanar on the primary $\{111\}$ but is nonetheless dissociated: the two SIA atoms of each dumbbell straddle their vacancy by $\pm \tfrac{1}{2}L\,\hat b$ along the modal $\langle110\rangle$, so the parallel half-vectors (vacancy~$\to$~SIA) sit on one $\{111\}$ layer and the anti-parallel half-vectors on an adjacent layer even when the vacancies themselves are coplanar. A secondary partition that projects the SIA atom positions onto the primary $\{111\}$ normal and bins them into adjacent integer layer indices recovers these Shockley dissociations whose two partial cores have glided onto adjacent $\{111\}$ layers while the dumbbell--vacancy line remains in the original plane.

\subsection{Algorithm Summary}
\label{sec:algorithm}

\begin{algorithm}[H]
\caption{Savi-Bhransha}\label{alg:savi}
\begin{algorithmic}[1]
\Require Atomic coordinates, lattice parameters, crystal structure
\Ensure Loop topology, Burgers vector, habit plane, character, boundary/bulk separation

\State Identify dumbbell, displaced atom--vacancy, and vacancy-centred motifs from lattice-site occupancy
\State Assign each line primitive to a crystallographic direction family
\State Group point defects into clusters; merge collinear lines within each cluster

\For{each cluster}
    \State Build adjacency graph with parallel and ring edges
    \State Find connected components; classify morphological units
\EndFor

\For{each parallel (loop) component}
    \State Fit habit plane via weighted PCA on line centroids; match to low-index family
    \State Assign Burgers family by consensus over constituent line primitives; disambiguate state using habit and defect content
    \State Count same-family neighbors; classify boundary vs bulk lines
    \State Construct $\alpha$-shape loop perimeter; compute segment-wise edge/screw character
    \If{FCC and $\langle110\rangle$ family}
        \State Partition dumbbells across the $\{111\}$ habit normal; report perfect $\frac{1}{2}\langle110\rangle$ or dissociated $\frac{1}{6}\langle112\rangle$ partial pair as appropriate
    \EndIf
\EndFor

\State \Return Loop morphology and quantitative descriptors
\end{algorithmic}
\end{algorithm}

\subsection{Computational Considerations}
\label{sec:implementation}

The cost of the Savi-Bhransha pipeline decouples from the simulation-cell size after the initial defect-identification step. Defect identification itself is the only stage that touches every atom; line merging, graph construction, component classification, and morphological characterisation all operate solely on the $N_{\text{defect}}$ line primitives within an individual cluster. Because cascade simulation boxes are typically chosen large enough to suppress finite-size artefacts, the clusters they contain remain modest: a $100{\times}100{\times}100$ unit-cell box with $\sim$2\,M atoms produces clusters of at most $10$--$10^{2}$ line primitives even at high PKA energies, so per-cluster analysis is essentially free relative to the defect-identification pass. Material dependence enters only through lattice geometry---nearest-neighbour distances and symmetry families---and every threshold in the pipeline is scaled from these lattice quantities rather than chosen as a bare cut-off.

The dominant cost within each cluster is pairwise line-distance evaluation. We expose a single computational kernel with three back-ends that share the same vectorized formulation: a CPU back-end using array broadcasting for the pairwise computation, a parallel-CPU back-end that distributes the per-line neighbor search across cores via just-in-time compilation, and a CUDA back-end that runs the same kernel on the GPU for the largest clusters. The resulting adjacency graph is sparse---each line has at most four to eight parallel neighbors set by the lattice---so it contains $O(N_{\text{defect}})$ edges rather than $O(N_{\text{defect}}^2)$. Orientation families are carried as integer codes throughout, which keeps every inner loop free of string comparisons and amenable to vectorization.

Memory follows the same decoupling: only the per-cluster $O(N_{\text{defect}})$ line primitives and their sparse adjacency are held, in contrast to the $O(N_{\text{atoms}})$ connectivity data that mesh-based extraction must maintain over the entire cell. At the largest HCP system tested (172,M atoms), Savi-Bhransha completes the analysis in approximately 5,min using 24,GB, compared with 185,GB for DXA; the full benchmark is reported in Section~\ref{sec:results}.

\subsection{MD Cascade Datasets}
\label{sec:md_datasets}

The cascade trajectories analysed in Section~\ref{sec:results} were generated with LAMMPS~\cite{plimpton1995lammps,thompson2022lammps} for two distinct loading regimes: single-PKA cascades for surveying primary damage as a function of energy and crystal structure, and successive collision cascades (SCC) for tracking microstructural evolution under accumulated dose. Both regimes share the same per-cascade procedure---NPT equilibration at 300\,K and zero pressure, an NVE cascade phase with an adaptive timestep that keeps the fastest atom's displacement below $0.1$\,\AA\ per step, and Lindhard--Scharff electronic stopping treated as a frictional drag~\cite{lindhardscharff1961}---and use ZBL-stiffened pair interactions at short range~\cite{zbl1985}. Material dependence enters through the choice of interatomic potential and threshold displacement energy $E_d$.

\paragraph{Single-PKA cascades.}
A PKA is launched from the centre of a cubic, periodic supercell in a randomly chosen direction; many independent directions per energy are sampled to obtain stable averages, following the protocol of Warrier et~al.~\cite{warrier2015pka,warrier2024cascade}. The cascade is integrated for $6$--$10$\,ps in the NVE ensemble, after which the surviving Frenkel content is read off the relaxed configuration. We use this protocol for the BCC W energy sweep (10--150\,keV, DnD-BN~\cite{marinica2013w} and SNAP~\cite{wood2017snap} potentials, $E_d = 70$\,eV) and the HCP Zr energy sweep (10--125\,keV, Starikov--Smirnova ADP potential~\cite{starikov2021zrnb}, $E_d = 40$\,eV).

\paragraph{Successive collision cascades.}
Long-time damage accumulation is generated by chaining individual cascades: after each NVE cascade, the system is relaxed in NPT at 300\,K and zero pressure for $\sim$10\,ps, then a fresh PKA is selected from a randomly drawn lattice atom and launched along a fresh random direction; the cycle repeats until the desired dose is reached~\cite{bhardwaj2024scc}. The simulation cell remains periodic and unconstrained (no fixed boundary), so the lattice can shift collectively as material is displaced---a feature accommodated by the defect-identification step of Section~\ref{sec:defect_id}. Dose is reported in displacements per atom (dpa) using the standard NRT formula~\cite{norgett1975nrt}. Five independent trajectories per condition are integrated to expose stochastic spread. We apply this protocol to BCC W at 50\,keV (DnD-BN~\cite{marinica2013w} and SNAP~\cite{wood2017snap}, up to 0.2\,dpa) and to FCC FeNiCr at 20\,keV with the EAM potential of Bonny et~al.~\cite{bonny2013} ($E_d = 40$\,eV), accumulated to $7.205\times10^{-3}\,dpa$ over the 250-cascade trajectory. For the FCC series, the dose uses the standard NRT displacement count $\nu_{\mathrm{NRT}}=0.8E/(2E_d)$.

\section{Results}
\label{sec:results}

We evaluate Savi-Bhransha on three cascade datasets spanning the BCC, HCP, and
FCC lattices: single-PKA damage in BCC W and HCP Zr, and successive cascades up
to 0.2\,dpa in BCC W and FCC FeNiCr. Together these cover sparse isolated
defects, dense mixed defect populations, compact debris, vacancy-rich regions,
and stacking-fault-related FCC morphologies. The section first reports
material-specific loop morphology---Burgers-vector and habit-plane populations,
loop size, vacancy/interstitial sign, boundary--bulk separation, and behavior
in complex cascade debris---then compares Savi-Bhransha with DXA on loop count
and total length, and finally summarizes timing and memory.

\subsection{Irradiation in BCC W}

For BCC W we use two complementary cascade datasets: (i) a 100-cascade single-PKA energy sweep (10--150~keV, two potentials, 10 replicas per energy--potential combination) that fixes the morphology of primary damage as a function of PKA energy, and (ii) a successive-cascade trajectory at 50~keV up to 0.2\,dpa that follows the same morphology under accumulated dose. We discuss the two in turn.

\subsubsection*{Single-PKA energy sweep}

\begin{figure*}[!ht]
\centering
\includegraphics[width=0.8\textwidth]{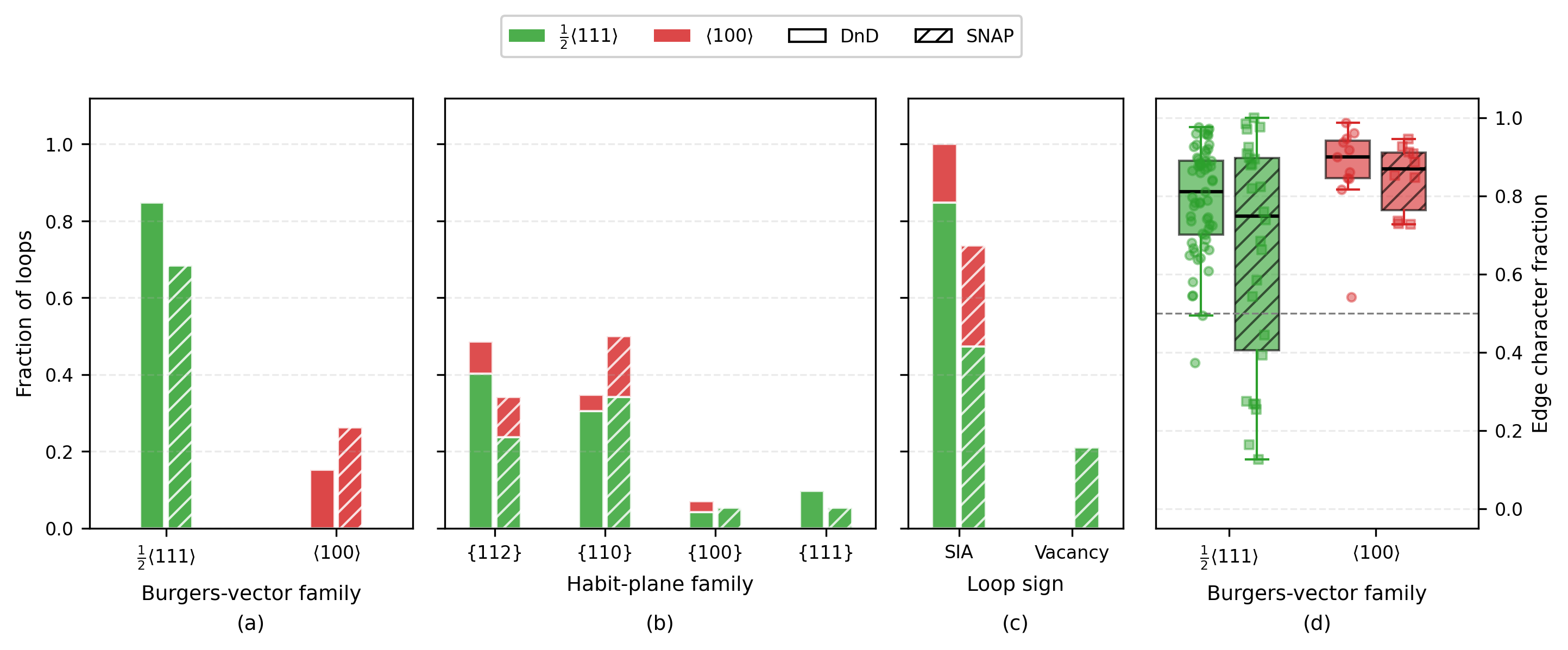}
\caption{Aggregated loop morphology in BCC W from 10--150~keV PKA simulations (DnD and SNAP potentials, 10 replicas per energy--potential combination, 100 cascades total).
(a)~Burgers vector population fractions; solid bars: DnD, hatched bars: SNAP.
(b)~Habit plane family distribution per Burgers vector type.
(c)~Interstitial vs vacancy loop type fractions.
(d)~Edge character fraction per Burgers vector type and potential; box plots show median, interquartile range, and outliers; dashed line marks the edge/screw boundary at $f_\text{edge}=0.5$.}
\label{fig:w_bcc_summary}
\end{figure*}

We analyze 100 single collision-cascade simulations in BCC tungsten at 10--150~keV PKA energies using the DnD embedded-atom potential and the SNAP machine-learning potential, with 10 cascades per energy--potential combination. The loop count rises from near zero at 10~keV to ${\sim}3$--4 significant loops per cascade at 150~keV; DnD consistently produces more loops than SNAP at equivalent energies, reflecting potential-level differences in cascade morphology~\cite{warrier2024cascade}. Figure~\ref{fig:w_bcc_summary} summarizes the aggregated loop morphology across all energies.

\paragraph{Burgers vector.}
$\frac{1}{2}\langle111\rangle$ loops dominate (panel~a), comprising ${\sim}85\%$ of DnD loops and ${\sim}68\%$ of SNAP loops across all energies. $\langle100\rangle$ loops make up the remainder and appear predominantly at $\geq$100~keV, consistent with the established trend that their formation requires sufficient cascade energy~\cite{setyawan2015,setyawan2023}; SNAP produces a larger $\langle100\rangle$ fraction than DnD, in line with prior multi-potential comparisons~\cite{warrier2024cascade,bhardwaj2022loop}.

\paragraph{Habit planes.}
$\{112\}$ and $\{110\}$ planes dominate (panel~b), with $\{110\}$ preferentially associated with $\frac{1}{2}\langle111\rangle$ loops and $\{100\}$ with $\langle100\rangle$ loops, as expected from crystallographic considerations.

\paragraph{Vacancy vs interstitial loops.}
The vast majority of loops are interstitial (panel~c); vacancy loops appear only with the SNAP potential, and only at PKA energies of 100~keV and above.

\paragraph{Edge/screw character.}
Loops are predominantly edge in character (panel~d). The broader distribution for $\frac{1}{2}\langle111\rangle$ loops under SNAP reflects greater morphological diversity in this population.

\paragraph{Loop size.}
Diameters are predominantly sub-nanometer, with mean circular diameter ${\sim}1$~nm at 100--150~keV; most loops fall below the typical TEM visibility threshold of $\sim$2~nm.

\subsubsection*{Successive-cascade dose evolution}

\begin{table*}[ht]
\centering
\footnotesize
\caption{Comparison of MD-derived observables with experimental literature at comparable doses. MD values are from the DnD and SNAP successive-cascade trajectories at 50~keV (mean over five independent trials), reported as DnD\,/\,SNAP. $\Sigma$: areal density, Method~B, $f_\text{vis}=2/3$, $\geq$2.0~nm. All experimental data are for W self-ion irradiation at room temperature. Bold entries highlight near-quantitative agreement: the boundary-defect concentration brackets the TGS measurement, $\Sigma$ at 0.01\,dpa matches the Yi et~al.\ TEM value, and the SNAP $\frac{1}{2}\langle111\rangle$ fraction at 0.20\,dpa matches the Yi et~al.\ room-temperature value. The PAS comparison is qualitative: Hollingsworth et~al.\ report saturation of the irradiation-defect density over the 0.085--0.425\,dpa window rather than an absolute concentration.}
\label{tab:dpa_exp_comparison}
\setlength{\tabcolsep}{6pt}
\begin{tabular}{llll}
\toprule
Observable & MD (this work) & Experiment & Reference \\
\midrule
Defect conc.\ at 0.20~dpa            & $7.3 / 5.7 \times10^{-3}$ & saturates 0.085--0.425~dpa & Hollingsworth et~al.\ (PAS)~\cite{amin2020pas} \\
Defect conc.\ at 0.1~dpa (total)     & $4.7 / 3.6 \times10^{-3}$ & $2.25\times10^{-3}$         & TGS, Reza et~al.\ (2020)~\cite{reza2020tgsthermal} \\
Defect conc.\ at 0.1~dpa (boundary)  & $\mathbf{1.8 / 2.7 \times10^{-3}}$ & $\mathbf{2.25\times10^{-3}}$ & TGS, Reza et~al.\ (2020)~\cite{reza2020tgsthermal} \\
$\Sigma$ at 0.01~dpa                 & $\mathbf{1.6 / 1.3 \times10^{15}}$~m$^{-2}$ & $\mathbf{\sim1\times10^{15}}$~m$^{-2}$ & Yi et~al.\ (2016)~\cite{yi2016} \\
$\Sigma$ at 0.2~dpa                  & $14 / 7.9 \times10^{15}$~m$^{-2}$ & $\sim4\times10^{15}$~m$^{-2}$ & Yi et~al.\ (2016)~\cite{yi2016} \\
Mean $d$ at 0.05~dpa                 & $1.56 / 1.28$~nm          & $7.3\pm2.5$~nm at 0.04~dpa  & Wieluńska et~al.\ (2022)~\cite{wielunska2022} \\
$\frac{1}{2}\langle111\rangle$ frac.\ at 0.20~dpa & $68 / \mathbf{77}$\%      & $\mathbf{\sim75\%}$ at RT            & Yi et~al.\ (2016)~\cite{yi2016} \\
\bottomrule
\end{tabular}
\end{table*}

\begin{figure*}[!ht]
\centering
\includegraphics[width=0.9\textwidth]{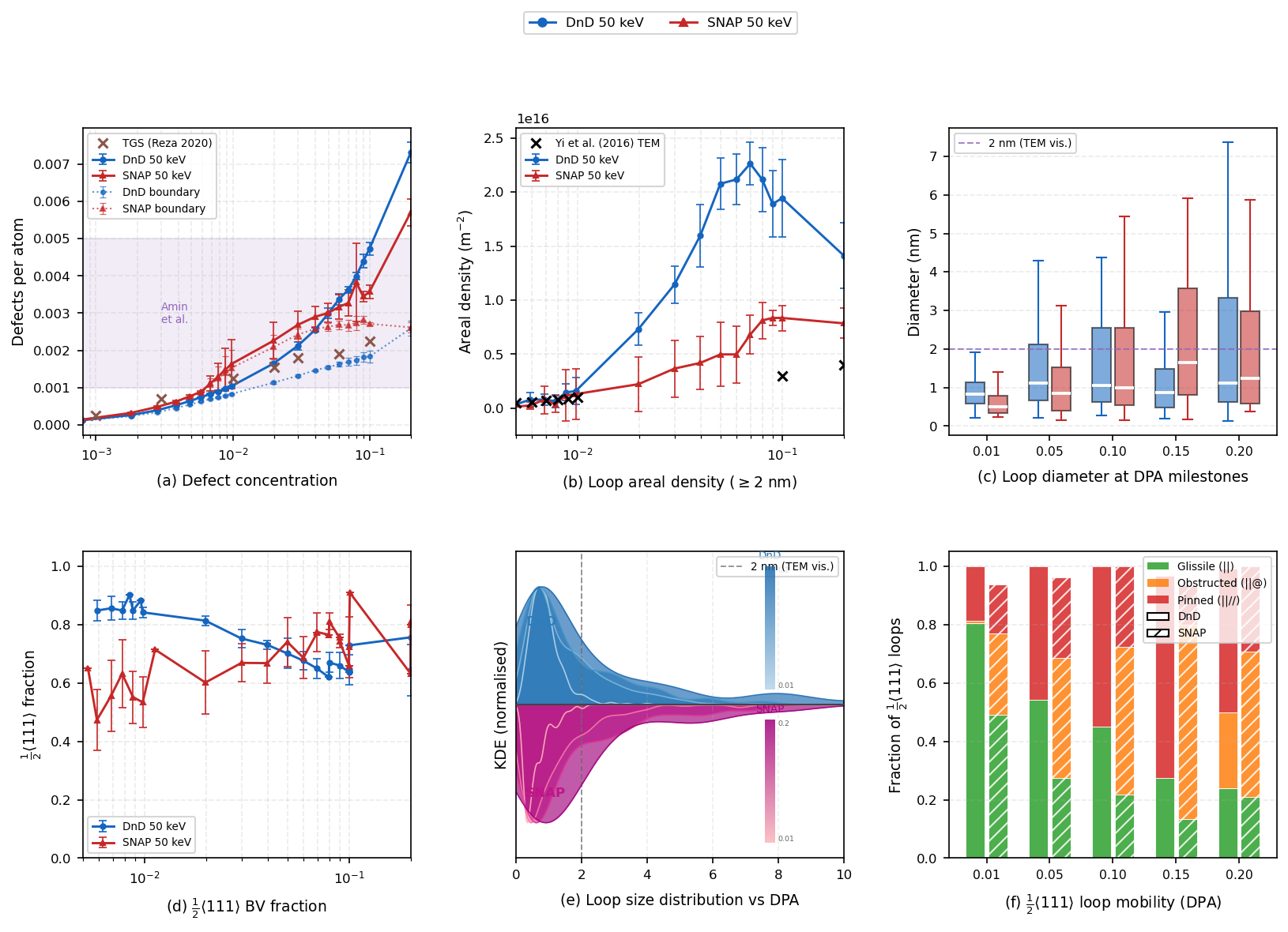}
\caption{Evolution of defect and loop metrics under successive 50\,keV W cascades (DnD and SNAP potentials, five trials per potential). Error bars indicate $\pm1\sigma$ inter-trial spread.
(a)~Defect concentration; Amin et~al.\ PAS saturation band and TGS data from Reza et~al.\ (2020) shown for reference. The MD \emph{boundary-defect} populations (dotted lines) bracket the TGS curve, supporting a boundary-driven interpretation of the TGS signal.
(b)~TEM-visible loop areal density ($\Sigma = N_{\geq2\,\text{nm}}/A_\text{proj}\times f_\text{vis}$, $f_\text{vis}=2/3$); $\times$~TEM data from Yi et~al.\ (2016). MD and TEM agree closely at 0.01\,dpa and diverge at higher dose.
(c)~Loop circular diameter at DPA milestones; dashed line marks the 2\,nm TEM visibility threshold.
(d)~$\frac{1}{2}\langle111\rangle$ Burgers vector fraction vs dose; SNAP at 0.20\,dpa matches the $\sim$75\% room-temperature value of Yi et~al.~\cite{yi2016}.
(e)~Normalised loop size KDE at 0.01 and 0.2\,dpa for each potential.
(f)~$\frac{1}{2}\langle111\rangle$ loop mobility fractions: glissile ($\|$), obstructed ($\|\!@$), and pinned ($\|/\!\!/$).}
\label{fig:dpa_evolution}
\end{figure*}

Figure~\ref{fig:dpa_evolution} presents the evolution of defect and loop metrics under successive 50\,keV cascades for the DnD and SNAP potentials, accumulated up to 0.2\,dpa across five independent trial trajectories.

\paragraph{Defect accumulation.}
Defect concentration rises steeply through 0.05\,dpa for both potentials (panel~a), with DnD reaching $4.7\times10^{-3}$ at 0.1\,dpa and $7.3\times10^{-3}$ at 0.2\,dpa; SNAP accumulates slightly more slowly ($3.6\times10^{-3}$ and $5.7\times10^{-3}$ at the same doses). The growth slows after 0.1\,dpa, consistent both with the PAS defect-density saturation that Hollingsworth et~al.~\cite{amin2020pas} report over the 0.085--0.425\,dpa window in self-ion irradiated W and with the $\sim$55\% thermal-diffusivity drop measured by TGS at that dose~\cite{reza2020tgsthermal}. The boundary-defect populations $1.8\times10^{-3}$ (DnD) and $2.7\times10^{-3}$ (SNAP) at 0.1\,dpa bracket the TGS measurement of $2.25\times10^{-3}$~\cite{reza2020tgsthermal}, reinforcing the interpretation that TGS is sensitive primarily to the boundary population rather than to the total defect content. Beyond 0.1\,dpa the gap between total and boundary populations widens as loops overlap and grow by absorption---reflected in the falling loop number density and rising mean diameter (panels~b,~c). The total defect concentration continues to climb while the boundary fraction saturates, consistent with the corresponding plateau in the TGS signal.

\paragraph{Loop areal density.}
Loop areal density $\Sigma = N_{\geq2\,\text{nm}}/A_\text{proj}\times f_\text{vis}$ (panel~b), where $A_\text{proj} = L_xL_y$ and $f_\text{vis}=2/3$ accounts for diffraction-invisible loops under the two standard $g$-vectors in BCC~W, peaks near $2\times10^{16}$\,m$^{-2}$ for DnD around 0.05--0.10\,dpa before relaxing toward $\sim$$1.4\times10^{16}$\,m$^{-2}$ at 0.2\,dpa; SNAP rises more slowly to $\sim$$8\times10^{15}$\,m$^{-2}$ at 0.10--0.20\,dpa. \textbf{At very low dose (0.01\,dpa) both potentials match the Yi et~al.\ TEM measurement closely}: $1.6\times10^{15}$ (DnD) and $1.3\times10^{15}$ (SNAP) versus $\sim$$1\times10^{15}$\,m$^{-2}$~\cite{yi2016}. By 0.1--0.2\,dpa the MD values run 2--4$\times$ above the same TEM band, suggesting a large sub-threshold loop population that is captured by Savi-Bhransha but partially missed by conventional microscopy as the population coarsens.

\paragraph{Loop size and hardening implications.}
Mean loop diameters grow from $\sim$0.9\,nm at 0.01\,dpa to $2.4\pm0.3$\,nm (DnD) and $3.0\pm0.4$\,nm (SNAP) at 0.2\,dpa (panel~c). The broad size distributions at high dose (panel~e) reflect a few large merged clusters coexisting with a majority of sub-threshold loops. By 0.1\,dpa the median DnD loop has approached the 2\,nm TEM visibility threshold, placing it in the partial-absorption regime identified by Lin et~al.~\cite{lin2024cpfem}; SNAP loops follow a similar trajectory but with a broader spread.

\paragraph{Burgers vector evolution.}
The two potentials show contrasting dose dependence (panel~d). DnD starts $\frac{1}{2}\langle111\rangle$-dominated ($85\pm2\%$ at 0.01\,dpa), and the $\langle100\rangle$ fraction grows from $15\%$ to $\sim$30\% by 0.05\,dpa where it stabilises ($29\pm7\%$ at 0.20\,dpa). SNAP begins with a much lower $\frac{1}{2}\langle111\rangle$ fraction ($57\pm11\%$ at 0.01\,dpa) and a correspondingly large $\langle100\rangle$ population ($41\pm10\%$); the $\frac{1}{2}\langle111\rangle$ share rises monotonically to $77\pm8\%$ at 0.20\,dpa as $\langle100\rangle$ loops are preferentially absorbed or converted, suggesting potential-dependent cross-slip and annihilation pathways~\cite{warrier2024cascade}. \textbf{The SNAP value at 0.20\,dpa lies within the experimental uncertainty of the $\sim$75\% room-temperature $\frac{1}{2}\langle111\rangle$ fraction reported by Yi et~al.~\cite{yi2016}}, while DnD remains slightly lower at the same dose.

\paragraph{Loop mobility.}
Panel~(f) decomposes $\frac{1}{2}\langle111\rangle$ loops by mobility class---glissile ($\|$, isolated parallel component), obstructed ($\|\!@$, parallel component entangled with a ring), and pinned ($\|/\!\!/$, parallel component in a multi-component complex). At 0.01\,dpa DnD loops are overwhelmingly glissile ($82\%$, with the remainder pinned), whereas SNAP already shows a sizeable obstructed population ($30\%$) on top of $48\%$ glissile and $15\%$ pinned. As dose accumulates the two potentials diverge: DnD shifts toward a \emph{pinned}-dominated state ($55\%$ pinned, $29\%$ glissile, $16\%$ obstructed at 0.20\,dpa), while SNAP becomes \emph{obstructed}-dominated ($48\%$ obstructed, $25\%$ pinned, $24\%$ glissile at the same dose). The persistent obstructed fraction in SNAP directly reflects the higher prevalence of C15-like ring structures in SNAP cascades acting as entanglement centres for otherwise mobile $\frac{1}{2}\langle111\rangle$ loops---a mechanism with direct implications for cascade-overlap hardening models.

\subsection{Collision Cascades in HCP Zr}
\label{subsec:zr_hcp}

\begin{figure*}[!ht]
\centering
\includegraphics[width=0.8\textwidth]{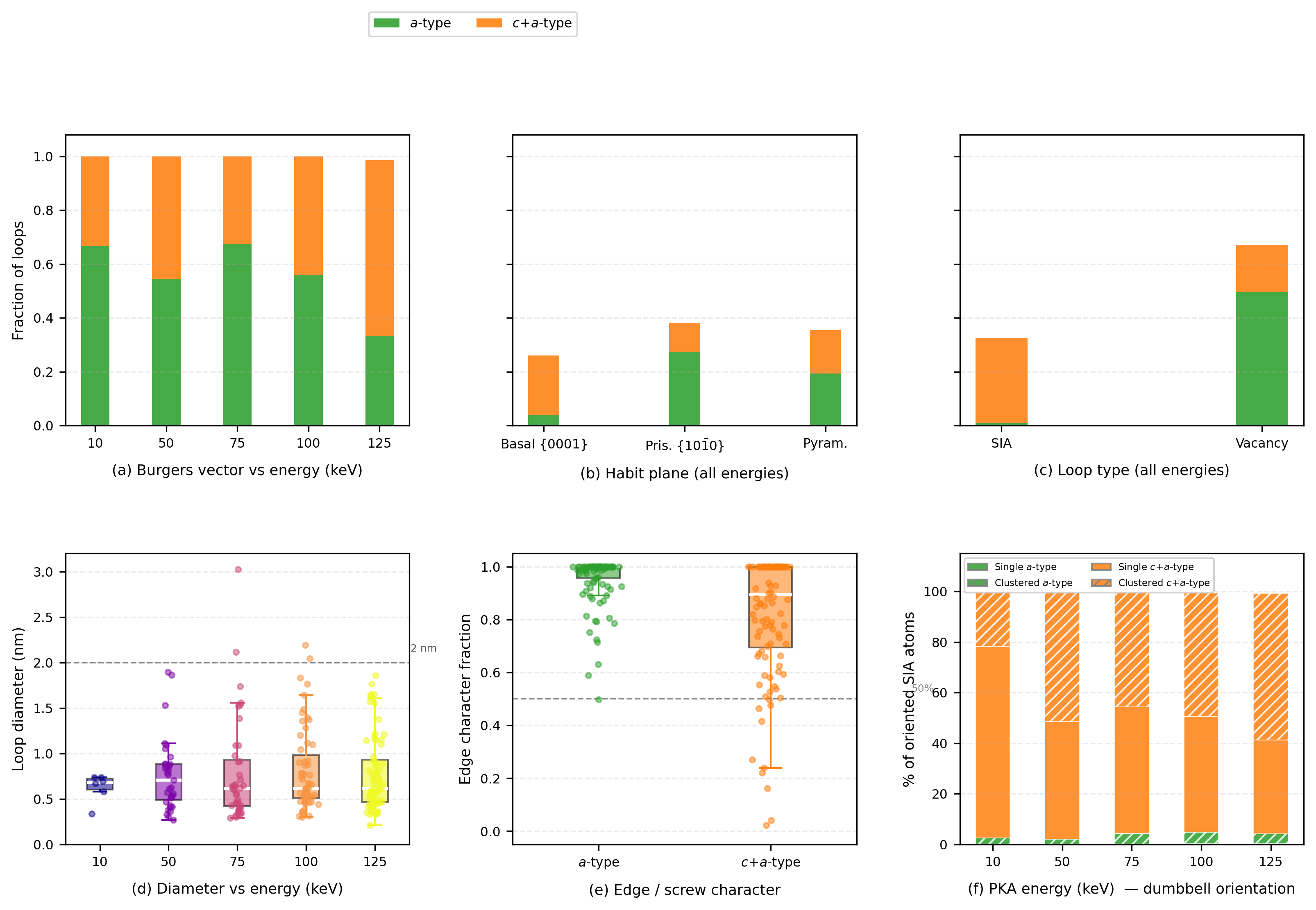}
\caption{Single-cascade loop morphology in HCP Zr from 10--125~keV PKA simulations (ADP potential, 5 replicas each).
(a)~Burgers vector fractions ($\langle a\rangle$: green; $\langle c\!+\!a\rangle$: orange) per PKA energy; $\langle a\rangle$ dominates at 10--100~keV, $\langle c\!+\!a\rangle$ at 125~keV.
(b)~Habit-plane family fractions aggregated over all energies, coloured by Burgers vector type.
(c)~Loop type (SIA / vacancy) fractions aggregated over all energies, coloured by Burgers vector type.
(d)~Loop circular diameter per energy; dashed line marks the 2~nm TEM visibility threshold.
(e)~Edge character fraction by Burgers vector type (all energies pooled).}
\label{fig:zr_hcp_paper}
\end{figure*}

We apply Savi-Bhransha to 25 single-cascade simulations in hexagonal close-packed (HCP) zirconium spanning 10--125~keV using the Starikov--Smirnova ADP interatomic potential~\cite{starikov2021zrnb}, with five replicas per energy. Figure~\ref{fig:zr_hcp_paper} summarizes the extracted loop morphology.

\paragraph{Defect yield and loop formation.}
Surviving defect counts scale from $\sim$36 Frenkel pairs at 10~keV to $\sim$634 at 125~keV, consistent with the NRT model modified by a cascade efficiency of 0.4--0.6~\cite{nordlund2018}. The mean loop count per cascade grows correspondingly from 1.2 at 10~keV to 14.4 at 125~keV.

\paragraph{Burgers vector distribution.}
$\langle a\rangle$-type loops ($\frac{1}{3}\langle11\bar{2}0\rangle$) dominate at 10--100~keV (54--68\%), while at 125~keV $\langle c\!+\!a\rangle$-type loops ($\frac{1}{3}\langle11\bar{2}3\rangle$) take over the majority ($\sim$65\%; panel~a); $\langle c\rangle$-type loops remain negligible ($<$1\%) at all energies. The $\langle a\rangle$-dominance at moderate energies matches prior MD studies of HCP Zr cascade damage~\cite{wooding1997,woo1992point}, while the high-energy crossover plausibly reflects sub-cascade branching that generates localised displacement spikes aligned with pyramidal planes and activates $\langle c\!+\!a\rangle$ slip systems~\cite{wooding1997,stoller2002}. The accumulated-dose TEM microstructure, in contrast, is overwhelmingly dominated by $\langle a\rangle$ loops on prismatic planes~\cite{griffiths1988}; the difference reflects long-range SIA migration and the preferential growth of glissile $\langle a\rangle$ loops between cascades---processes absent from the single-PKA picture and which we return to under vacancy versus interstitial bias below.

\paragraph{Habit planes.}
Aggregated over all energies, prismatic $\{10\bar{1}0\}$ planes are the most common habit (38\%), followed by first-order pyramidal (35\%) and basal $\{0001\}$ (26\%; panel~b). Prismatic and pyramidal habits correlate with $\langle a\rangle$ and $\langle c\!+\!a\rangle$ Burgers vectors respectively, in agreement with the glide-plane assignments from diffraction-contrast TEM of in-reactor Zr~\cite{griffiths1988}.

\paragraph{Vacancy vs interstitial loops.}
Vacancy loops constitute 67\% of all loops (panel~c), indicating a pronounced vacancy-loop bias in this single-cascade ensemble~\cite{stoller2002,nordlund2018}. The bias is the expected single-cascade signature: vacancies aggregate in the dense cascade core while SIAs are ejected to the cooler periphery as smaller, more diffuse clusters~\cite{woo1992point,christiaen2019}. As noted under the Burgers vector discussion, this single-cascade picture inverts under accumulated dose, where interstitial loops grow preferentially under sustained flux through SIA mobility and the one-dimensional bias of glissile $\langle a\rangle$ loops~\cite{griffiths1988,christiaen2019}.

\paragraph{Loop size and TEM visibility.}
Mean circular diameters lie between 0.62~nm (10~keV) and 0.86~nm (50~keV) with no systematic growth at higher energies (panel~d): higher PKA energies generate more loops (1.2 to 14.4 per cascade, as above) rather than larger ones. Between 95\% and 100\% of loops at every energy fall below the 2~nm TEM visibility threshold, so conventional in-situ TEM captures only the tail of the true single-cascade size distribution~\cite{nordlund2018,woo1992point}.

\paragraph{Edge/screw character.}
Both $\langle a\rangle$ and $\langle c\!+\!a\rangle$ loops are overwhelmingly edge in character (mean edge fraction 0.89~$\pm$~0.19; panel~e), consistent with the prismatic and pyramidal glide geometries of HCP Zr in which the Burgers vector lies nearly perpendicular to the loop normal~\cite{griffiths1988}. The broader distribution for $\langle c\!+\!a\rangle$ loops reflects the greater geometric diversity of pyramidal slip systems relative to the single prismatic family.

\subsection{Successive Cascades in FCC FeNiCr}
\label{subsec:fcc_fenickr}

\begin{figure*}[!ht]
\centering
\includegraphics[width=0.8\textwidth]{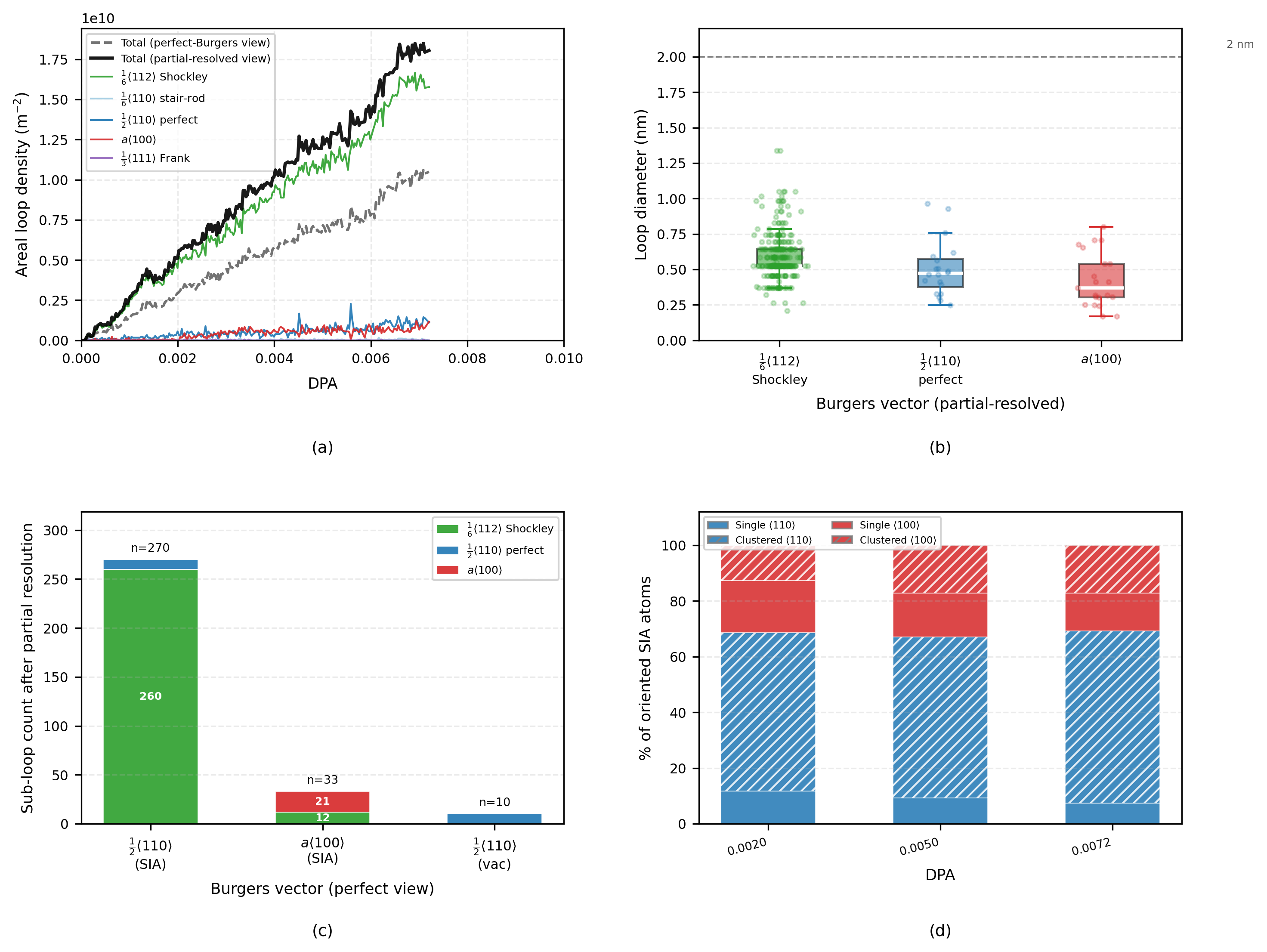}
\caption{Loop morphology in FCC FeNiCr ($120^3$ supercell, 250 successive 20~keV cascades, $7.205\times10^{-3}$~dpa by standard NRT).
(a)~Areal loop density versus dose: total perfect-Burgers loops (dashed black) and total partial-resolved sub-loops (solid black), with coloured traces giving the partial-resolved Burgers classes.
(b)~Sub-loop diameter at the terminal dose ($7.205\times10^{-3}$~dpa) for each partial-resolved Burgers class; dashed line marks the 2~nm TEM visibility threshold (stair-rods omitted---they are line-segment junctions with no closed-loop diameter).
(c)~Burgers transformation under partial resolution at the terminal dose: each column is a perfect-Burgers origin, stacked into the partial-resolved sub-loop classes it produces.
(d)~Dumbbell orientation fractions (solid = isolated, hatched = clustered) at representative doses; $\langle110\rangle$ dominates throughout.}
\label{fig:fcc_dpa}
\end{figure*}

The FCC FeNiCr alloy has a low stacking-fault energy, favouring Shockley-partial pairs that bound an intrinsic stacking fault~\cite{hirth1982}. Using Savi-Bhransha we report each surviving SIA cluster in two complementary views: a \emph{perfect-Burgers view} and a \emph{partial-resolved view} that decomposes $\langle110\rangle$-family clusters into Heidenreich--Shockley partials wherever the local dumbbell geometry supports it (Section~\ref{sec:shockley}). The dataset is 250 successive 20~keV cascades in a $120^3$-unit-cell supercell ($E_d = 40$~eV), accumulated to $7.205\times10^{-3}$~dpa under the standard NRT expression $\nu_{\mathrm{NRT}}=0.8E/(2E_d)$; vacancy loops and non-$\langle110\rangle$ clusters pass through both views unchanged. Figure~\ref{fig:fcc_dpa} summarises the loop morphology.

\paragraph{Topological loop population.}
Across the 250-cascade trajectory Savi-Bhransha identifies 20{,}693 significant loops (91\% SIA, 9\% vacancy). At the terminal analysed snapshot ($7.205\times10^{-3}$~dpa; PKA index 249) 189 loops survive, with the $\langle110\rangle$ family carrying $\sim$80\% of the population in the perfect-Burgers view (Fig.~\ref{fig:fcc_dpa}a, dashed trace), $a\langle100\rangle$ accounting for $\sim$14\%, and a few vacancy-labelled $\frac{1}{6}\langle112\rangle$ residuals. The $\langle110\rangle$ dominance is consistent with the 85--90\% reported by Osetsky et~al.~\cite{osetsky2018ni} for concentrated Ni alloys, and the contrast between glissile interstitial and faulted vacancy populations matches low-dose austenitic-steel data~\cite{jublotleclerc2016,lu2017ris}. Edge character is high throughout ($0.82\pm0.15$), agreeing with the $\geq$0.85 reported for interstitial loops in austenitic steels~\cite{edwards2010loops,stoller2002}.

\paragraph{Heidenreich--Shockley dissociation.}
The partial-resolved view recovers the dissociated state at the population level. Of the 17{,}380 SIA clusters analysed across the trajectory, 91\% carry a Shockley-pair signature via either the vacancy-lattice-site partition (64.7\%: 48.7\% two-band narrow-SF plus 16.0\% three-band resolved-SF) or the SIA-atom secondary partition (26.4\%; clusters whose vacancy lattice sites are coplanar but whose SIA atoms straddle two adjacent $\{111\}$ layers, Section~\ref{sec:shockley}). Only 7.7\% remain genuinely compact at the perfect $\frac{1}{2}\langle110\rangle$ label. The resulting sub-loop spectrum is overwhelmingly $\frac{1}{6}\langle112\rangle$ Shockley partials (93\%; Fig.~\ref{fig:fcc_dpa}a, green), with the $a\langle100\rangle$ channel at 5\% and residual undissociated $\frac{1}{2}\langle110\rangle$ cores at 2\%; genuine multi-plane $\frac{1}{6}\langle110\rangle$ stair-rod junctions account for only $\sim$0.3\%, giving a stair-rod-to-Shockley ratio well below 1:200. This near-complete dissociation of the $\langle110\rangle$-family SIA population is the qualitative outcome expected for the low stacking-fault-energy FeNiCr system~\cite{hirth1982,zhang2017sfe,zhang2019localsfe} and is consistent with the broader MD picture of faulted interstitial clusters in concentrated Ni-based alloys~\cite{beland2016}. The rarity of true stair-rods at this dose is also expected on theoretical grounds, since a sub-nm loop cannot easily span two $\{111\}$ glide planes~\cite{hirth1982,chen2024morph,osetsky2018ni}; partial-resolved sub-loops collapse onto $\{111\}$ as required by Heidenreich--Shockley crystallography~\cite{kaoumi2021stem}.

\paragraph{Sub-nm regime.}
All loops, in either view, remain below the 2~nm TEM visibility threshold (Fig.~\ref{fig:fcc_dpa}b): the largest single loop reaches 1.86~nm and Shockley partials average $\sim$0.5--0.7~nm, placing the population in the sub-nm ``black-spot'' damage regime~\cite{edwards2010loops,kaoumi2021stem}. The dominant single-dumbbell orientation is $\langle110\rangle$ throughout (Fig.~\ref{fig:fcc_dpa}d), matching the predicted ground-state SIA configuration in Ni, Cu, $\gamma$-Fe, and FeNiCr~\cite{nordlund2018,beland2016}. The persistence of dense sub-nm clusters at this dose---few loops above the TEM threshold, no surviving Frank loops, a high SIA-to-vacancy ratio---is consistent with delayed loop coarsening and high defect-cluster density reported under irradiation in concentrated Fe--Ni--Cr-family alloys~\cite{lu2017ris,zhao2023cssa}.

\subsection{Comparison with DXA}

In this section we present qualitative and quantitative comparisons between Savi-Bhransha and the widely used Dislocation Extraction Algorithm (DXA)~\cite{stukowski2010dxa} as implemented in OVITO~\cite{stukowski2010ovito}, drawing on the single-cascade BCC~W and HCP~Zr datasets and on a representative full-cell snapshot from the FCC~FeNiCr trajectory. DXA, which has become the de-facto standard for recovering dislocation networks from atomistic configurations, is taken as the reference; the discussion focuses on where the two views agree, where they differ, and what the differences imply for the choice of observables in cascade-microstructure analysis. Both algorithms operate directly on molecular-dynamics snapshots in which the atoms vibrate continuously about their nominal positions, so the comparison is inherently sensitive to snapshot-to-snapshot fluctuations and to the parameter choices of either method---a point we return to repeatedly below.

\subsubsection{Qualitative comparison}
\label{subsubsec:qualitative_dxa}

Across the three datasets, the loop populations resolved by Savi-Bhransha and DXA are in broad overall agreement: the major Burgers-vector families are recovered consistently by both methods, and the two reconstructions overlay closely in position and extent. The qualitative comparison below therefore concentrates on the minority of cases in which the two views diverge and on the systematic origin of those divergences. The DXA outputs shown were produced with the same default parameters across all snapshots; with different parameter choices, or even on adjacent frames of the same trajectory, individual loops can be returned as closed or as segmented, and occasional loops can appear or disappear as small atomic rearrangements move the configuration across the underlying mesh-construction threshold. The figures below are therefore not an exhaustive survey but a representative collection of edge cases that illustrate how each algorithm handles defect contexts which fall outside the simple isolated-loop picture.

\begin{figure*}[!ht]
\centering
\includegraphics[width=0.95\textwidth]{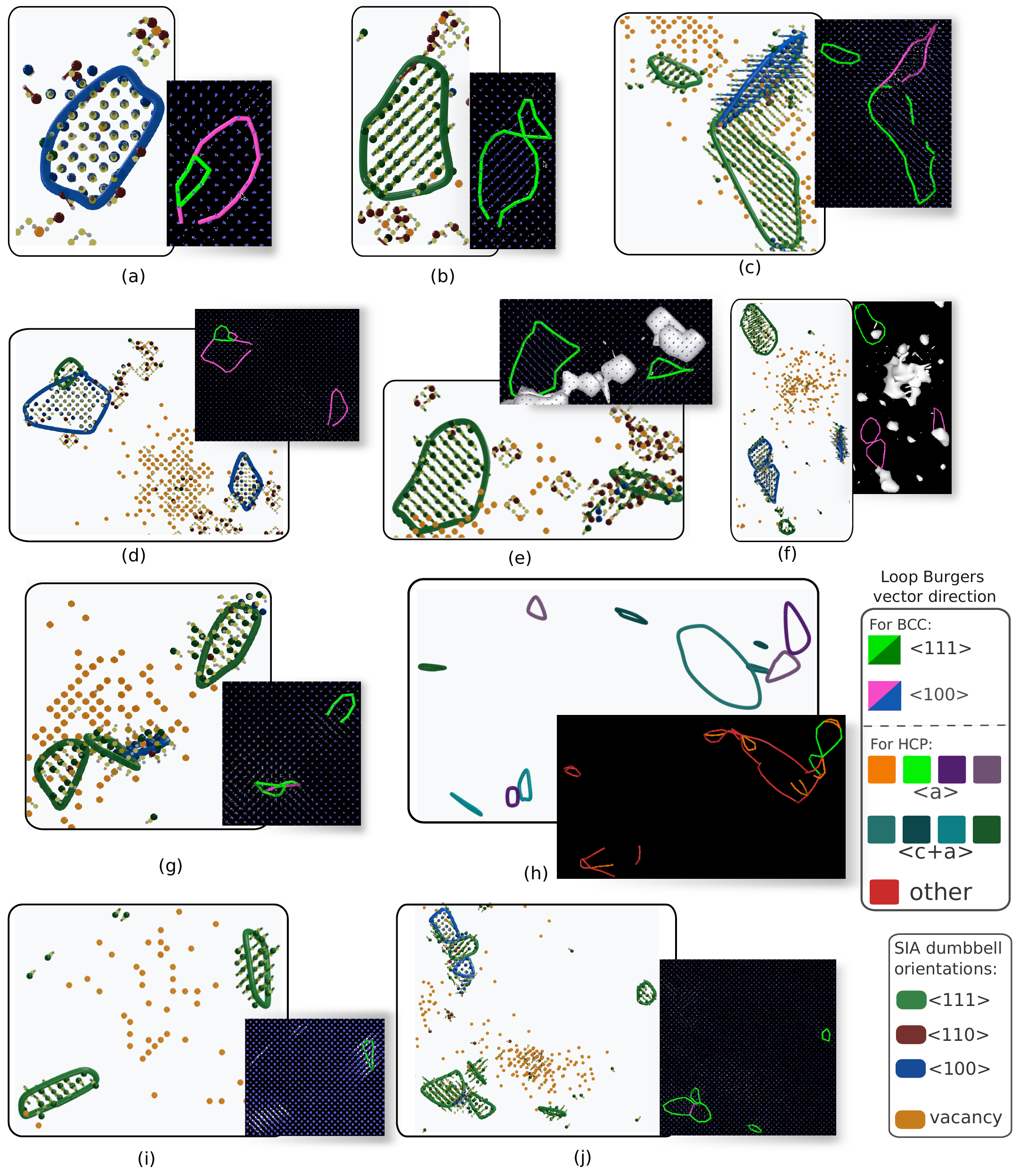}
\caption{Representative qualitative comparisons between Savi-Bhransha and DXA on the same atomistic configurations. In each panel the main image shows the Savi-Bhransha reconstruction with the constituent line primitives visible and the inset shows the DXA output (loops and segments coloured by Burgers vector; legend on the right). Panels (a)--(f) are BCC~W snapshots that illustrate loop fragmentation in the presence of proximate vacancies, point defects, $\langle100\rangle$-perimeter variability, or adjacent C15-like clusters; in (e) and (f) the DXA surface-mesh blobs are retained in the visualisation. Panels (g)--(j) illustrate the missing-loop failure mode: (g) and (i) are BCC~W snapshots and (h) and (j) are HCP~Zr snapshots in which one or more loops resolved by Savi-Bhransha are absent from the DXA output or are returned only as a partial trace; (h) additionally shows the heavy overlapping segmentation typical of many HCP~Zr cascades, with several segments carrying an ``unknown'' Burgers-vector label (red) that Savi-Bhransha identifies as $\langle c\!+\!a \rangle$.}
\label{fig:dxa_broken}
\end{figure*}

A recurrent source of disagreement is the splitting of a single physical loop into multiple DXA segments when the loop sits in a non-trivial defect environment. Figure~\ref{fig:dxa_broken} collects representative cases from the BCC~W dataset in which Savi-Bhransha and DXA agree on the positions, Burgers vectors, and habit-plane orientations of the loops but differ on the loop count or on the loop-versus-segment topology of individual objects. In some panels DXA splits a loop into two or more open segments in the presence of proximate vacancies, adjacent C15-like clusters, or isolated point defects on the loop perimeter; in others it misses one or more of the constituent components entirely. A precise mechanistic origin for the fragmentation is difficult to assign, and the behaviour appears to be a consequence of the local mesh construction and of the resulting Burgers-circuit analysis rather than of any specific morphological feature. The $\langle100\rangle$ loops in W are known to carry a variable number of dumbbells on their perimeter~\cite{bhardwaj2022loop}, and in panels (a), (c), (d), and (f) this variability appears to be one immediate reason that several $\langle100\rangle$ loops are returned as segments rather than as closed objects. In (a) the loop breaks on one side, while on the other side the $\langle111\rangle$ dumbbells are returned as a separate tiny $\langle111\rangle$ sub-loop. In (b) the proximity of a C15-like cluster makes the $\langle111\rangle$ loop break at one end and twist at the other. In (c) a nearby vacancy and a connected $\langle100\rangle$ loop together fragment the $\langle111\rangle$ loop into three open segments. In (d) a C15-like cluster fragments the loop on one side while the opposite side remains intact. In panels (e) and (f) we retain the mesh defect blobs produced by DXA in the OVITO visualisation: in (e) some C15-like defects are recovered as blobs while others are missed entirely, and the loops attached to those clusters are returned as pure $\langle111\rangle$ loops with no associated adjunct---here the presence of mesh blobs does not by itself fragment the loop. In (f), by contrast, the lower-left loop is split into segments by small nearby $\langle111\rangle$ clusters that appear as mesh blobs, the other two clusters carry blobs that originate from neighbouring point defects, and the $\langle100\rangle$ loop is again segmented. Mesh blobs do not by construction identify the defect content they enclose, whereas Savi-Bhransha returns the underlying line primitives directly and provides the additional information needed to interpret loop--cluster transition and interaction mechanisms~\cite{bhardwaj2022loop, CHARITHADURGA2026114658}. Non-dislocation defects more generally appear as mesh blobs in some configurations and are dropped from the analysis in others.

Panels (g)--(j) illustrate the complementary failure mode in which DXA either does not report a loop that Savi-Bhransha recovers on the same configuration, or returns only a partial trace of it. In (g) the lower-left component of a composite BCC~W cluster is absent from the DXA analysis while the upper loop is returned as a single open curved segment. Panel~(h) shows the heavy segmentation typical of many HCP~Zr cascades: a small loop population is reported as a much larger number of overlapping segments, several of which carry an ``unknown'' Burgers-vector label (red); the corresponding objects are identified as $\langle c\!+\!a \rangle$ by Savi-Bhransha. The segment-to-loop ratio in such cascades can exceed the loop count by an order of magnitude or more, which is the immediate origin of the count-statistics behaviour quantified in the next subsection. In~(i) only one of two comparably sized loops is recovered by DXA, and in~(j) a composite loop arrangement at the top of the snapshot is missed while smaller and comparably sized loops elsewhere in the cell are recovered.

Composite and pinned multi-component complexes are particularly prone to partial reporting in DXA: when one or more constituents of a multi-component arrangement are not recognised, the remaining glissile component is returned as if it were an isolated, unobstructed loop and the pinned context is masked. Because Savi-Bhransha treats each line primitive as an explicit graph node, the lower-level defect content and internal morphology are preserved and the multi-component context is retained alongside the loop label; the user can then choose, for instance, to flag an entangled loop as obstructed or segmented at the analysis stage rather than recovering that information through additional post-processing. The same loop can appear or disappear between adjacent MD frames as small atomic rearrangements move the configuration across the mesh-construction threshold, and this---together with the strong dependence of DXA on its mesh tolerances---limits the mechanistic significance that can be assigned to DXA segment counts themselves and skews loop-density and related statistics relative to a topology-stable reference.

The HCP~Zr dataset combines both failure modes. In a non-trivial fraction of cascades, DXA returns the dislocation content as many short pieces rather than as a small number of complete loops, with a sizeable share of the segment population carrying an ``unknown'' Burgers-vector label and several of the larger loops not recovered at all. In some cascades the segment count reaches the low hundreds while the closed-loop count returned alongside it is zero. If each segment is treated as a distinct loop downstream, density-type observables such as loop number density and the TEM-comparable areal loop density---which weight each physical loop once---will be inflated relative to the underlying population; HCP~Zr cascade dislocation analysis from MD is a notably challenging setting in this respect. The Savi-Bhransha output on the same configurations is comparatively cleaner: the loops are assembled into closed objects with stable Burgers-vector labels through the same defect environments. Total line length, by contrast, is far less sensitive to this fragmentation, and the two methods remain in good correspondence on that observable across the HCP set (Section~\ref{subsubsec:length_comparison}); we therefore treat total length as the more defensible cross-method scalar for HCP and rely on the loop-level Savi-Bhransha output for count- and morphology-based statistics.

\begin{figure*}[!ht]
\centering
\includegraphics[width=0.95\textwidth]{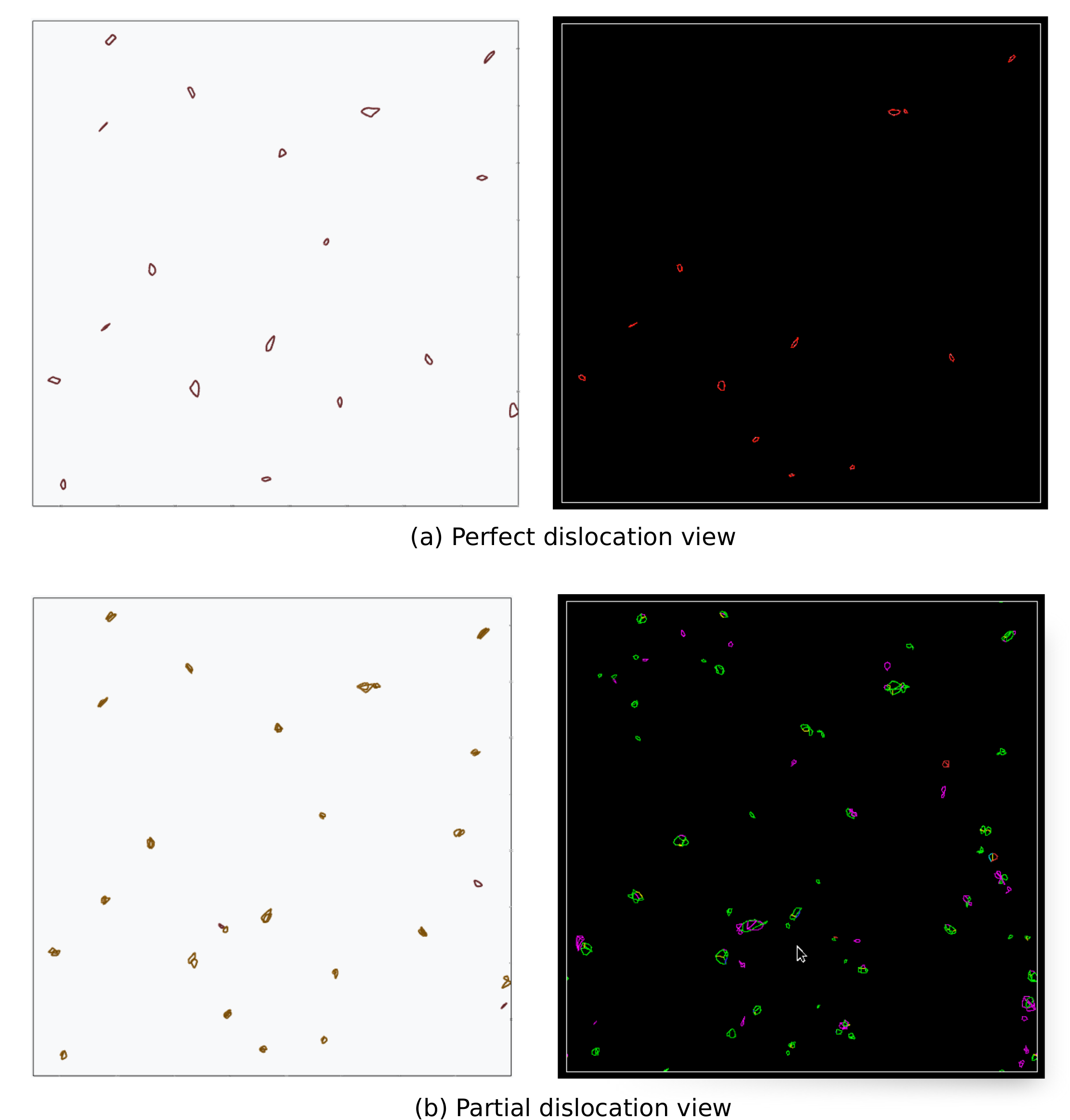}
\caption{Full-cell comparison on a representative FCC FeNiCr snapshot from the successive-cascade trajectory. (a) Perfect-Burgers view: Savi-Bhransha (left) and DXA (right) recover the same overall loop distribution, with Savi-Bhransha additionally returning a small number of loops not present in the DXA output. (b) Partial-resolved view: DXA reports many sub-resolution partial-dislocation fragments---including objects supported by only two or three dumbbells---whereas the Savi-Bhransha view applies a deterministic size cut-off (clusters of six dumbbells or fewer excluded) and retains the remaining well-resolved partial pairs.}
\label{fig:fcc_box_compare}
\end{figure*}

\paragraph{Full-cell view in FCC FeNiCr.}
Figure~\ref{fig:fcc_box_compare} extends the comparison to a complete simulation box from the FCC FeNiCr trajectory, in both the perfect-Burgers and partial-resolved views. At the cell scale the two methods recover the same overall loop distribution. In the perfect-Burgers view (panel~a), Savi-Bhransha additionally returns a small number of loops not present in the corresponding DXA picture, consistent with the segmentation behaviour discussed above. In the partial-resolved view (panel~b), DXA reports a substantially larger population of very small partial-dislocation fragments---including objects supported by only two or three dumbbells---while the Savi-Bhransha view retains only those clusters that exceed the size threshold imposed by the partial-dislocation analysis. The Savi-Bhransha threshold is a deterministic, user-tunable choice (clusters of six dumbbells or fewer are excluded from the partial analysis used here) rather than an algorithmic limit, and is set so that the reported partial-dislocation population is restricted to objects whose Burgers, habit, and dissociation geometry can be assigned with confidence.

Taken together, these qualitative comparisons indicate that the two methods, while broadly consistent on isolated well-formed loops, differ predominantly in how they handle loops embedded in non-trivial cascade debris, and that the resulting differences are systematic enough to be characterised rather than averaged out. They motivate the quantitative count- and length-based assessment that follows.

\subsubsection{Quantitative loop length and loop count comparison}
\label{subsubsec:length_comparison}

\begin{figure*}[!ht]
\centering
\includegraphics[width=0.95\textwidth]{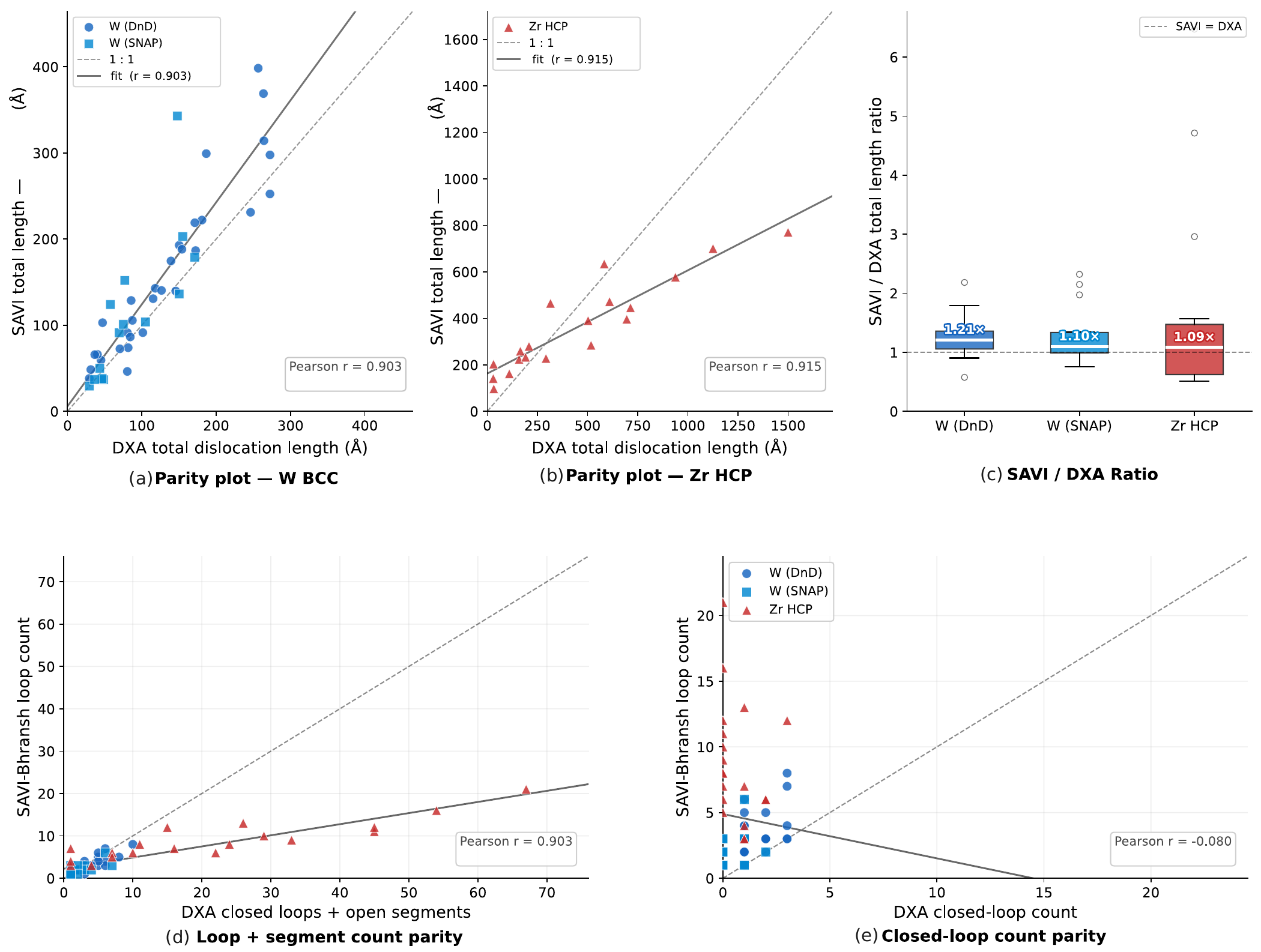}
\caption{Quantitative comparison between Savi-Bhransha and DXA across the filtered benchmark set of 64 cascades (W BCC DnD, W BCC SNAP, HCP Zr). (a) Parity plot of total dislocation length for BCC~W (DnD circles, SNAP squares); fitted line with Pearson $r=0.903$. (b) Parity plot of total dislocation length for HCP~Zr; $r=0.915$. (c) Box plots of the Savi-Bhransha/DXA total-length ratio by dataset, with median values labelled. (d) Loop count parity in which DXA closed loops and open segments are summed; $r=0.903$. (e) Loop count parity using DXA closed loops only; the correlation collapses ($r=-0.080$), with the majority of cascades sitting at or near DXA closed-loop count zero as many of the loops are identified as segments due to nearby defects.}

\label{fig:length_comparison}
\end{figure*}

Savi-Bhransha and DXA agree on total line length and loop count for the majority of cascades, with strong correlations in every subset (Fig.~\ref{fig:length_comparison}). For BCC~W (panel~a), the Savi-Bhransha total loop length is slightly higher than the DXA value with $r=0.903$; the excess is consistent with the non-deterministic loss of loops by DXA illustrated in Figs.~\ref{fig:dxa_broken}(i)~and~(j). For HCP~Zr (panel~b) the correlation remains strong ($r=0.915$), but the OLS slope deviates further from the 1:1 line. Two regimes are visible. At low PKA energies, where Zr cascades produce fewer and smaller loops, the Savi-Bhransha total is higher than DXA for the same reasons as in BCC. At high PKA energies, where defect density, size, and complexity grow rapidly, DXA returns highly fragmented overlapping segments---some carrying an ``unknown'' Burgers label---as shown in Fig.~\ref{fig:dxa_broken}(h); summing these segments inflates the DXA total length, while the Savi-Bhransha total tracks the underlying mostly-intact loop population. The median Savi-Bhransha/DXA length ratios are $1.21\times$ (W DnD), $1.10\times$ (W SNAP), and $1.09\times$ (Zr HCP), so the cross-method length difference is modest for normal cases with a small number of outliers in each subset (panel~c).

The loop-count comparison is shown in two complementary forms in panels~(d) and~(e). When DXA closed loops and open segments are summed (panel~d), the correlation with the Savi-Bhransha loop count is strong, though the DXA values are systematically higher in the BCC subsets because each fragmented physical loop contributes several segments. When DXA closed loops alone are used (panel~e), the correlation collapses ($r=-0.080$); the effect is most severe for the high-energy HCP~Zr cascades, where the DXA closed-loop count is small or zero even when Savi-Bhransha resolves several distinct loops. The high-energy HCP~Zr cascades contain overlapping loops, complex vacancy loops, and other debris. DXA often returns segments that cannot be assembled into a complete loop and terminate where the local dumbbell orientations no longer support a consensus; Savi-Bhransha therefore does not classify those segments as dislocation loops.

A post-processing pass that re-stitches DXA segments before counting can recover loop counts that are visually closer to the Savi-Bhransha values, and averaging over several consecutive snapshots, or analysing a locally minimised configuration rather than a dynamic one, can further reduce the frame-to-frame variability discussed above.

\subsubsection{Benchmark Timing and Memory}

\begin{figure*}[!ht]
\centering
\includegraphics[width=0.95\textwidth]{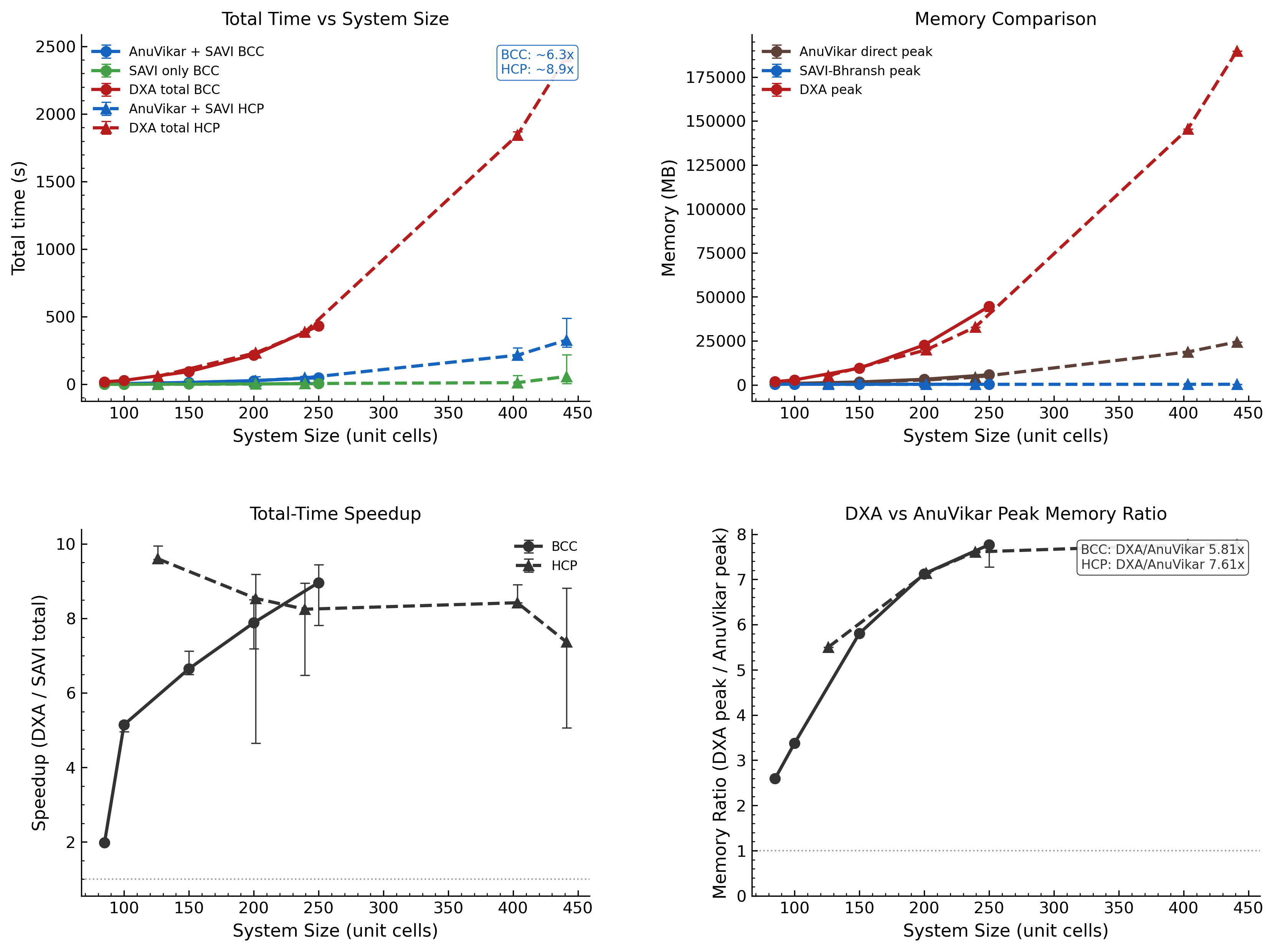}
\caption{Benchmark comparison of the Savi-Bhransha workflow (AnuVikar preprocessing plus graph analysis) and DXA across
W BCC (circles, solid) and Zr HCP (triangles, dashed) cascades as a function
of system size. Error bars show the interquartile range over cascade replicas at each energy level.
\textit{Top left}: total wall time.
\textit{Top right}: peak RSS memory, with the Savi-Bhransha graph-analysis
component (graph-analysis step only) shown separately to demonstrate its negligible
contribution.
\textit{Bottom left}: total-time speedup (DXA/Savi-Bhransha), reaching
up to $10.07\times$.
\textit{Bottom right}: DXA/AnuVikar peak-memory ratio, reaching
$7.77\times$ at the largest HCP system sizes.}
\label{fig:benchmark}
\end{figure*}

Figure~\ref{fig:benchmark} and Table~\ref{tab:benchmark_summary} summarise wall-time and peak-memory benchmarks across 50 BCC~W cascades (20--200\,keV PKA energy, 1.2--31\,M atoms) and 26 HCP~Zr cascades (10--125\,keV, 4--172\,M atoms). All timing and memory series are measured directly.

\paragraph{Timing.}
The median total wall time of the Savi-Bhransha pipeline is 14.24\,s for BCC and 62.66\,s for HCP, compared with 94.0\,s and 385.3\,s for DXA, corresponding to median speedups of $6.34\times$ and $8.86\times$, respectively. Speedups increase with system size: at the largest BCC system
(31\,M atoms, 200\,keV) the median speedup reaches $9.87\times$, and at the
largest HCP system (172\,M atoms, 125\,keV) it is $7.37\times$; the
single-run maximum across the full dataset is $10.07\times$.
The Savi-Bhransha graph-analysis component accounts for a small fraction of total
pipeline time---a median of 0.06\,s at BCC 50\,keV, growing to 5.25\,s at
200\,keV and 58\,s at HCP 125\,keV---confirming that the bottleneck is the
defect-identification step (AnuVikar / Avi), not the loop characterization.

\begin{table}[t]
\centering
\footnotesize
\setlength{\tabcolsep}{3pt}
\caption{Median benchmark metrics. Speedup = $t_{\text{DXA}}/t_{\text{SB}}$, where SB denotes the Savi-Bhransha workflow; mem ratio = DXA peak / AnuVikar peak RSS. BCC: 50 cascades (20--200\,keV); HCP: 26 cascades (10--125\,keV).}
\label{tab:benchmark_summary}
\begin{tabular}{lrrrr}
\toprule
System & $t_{\text{SB}}$\,(s) & $t_{\text{DXA}}$\,(s) & Speedup\,med & Speedup\,max \\
\midrule
W BCC  & 14.24 &  94.02 & $6.34\times$ & $9.87\times$ \\
Zr HCP & 62.66 & 385.32 & $8.86\times$ & $10.07\times$ \\
\bottomrule
\end{tabular}
\vspace{2mm}

\begin{tabular}{lrrrr}
\toprule
System & Avi (MB) & DXA (MB) & Ratio med & Ratio max \\
\midrule
W BCC  &  1629 &  9456 & $5.81\times$ & $7.77\times$ \\
Zr HCP &  4321 & 32873 & $7.61\times$ & $7.77\times$ \\
\bottomrule
\end{tabular}
\end{table}

\paragraph{Memory.}
The memory advantage is equally pronounced and grows with system size. The median DXA/AnuVikar peak-RSS ratio is $5.81\times$ for BCC and $7.61\times$ for HCP. At the largest tested system (172\,M-atom HCP, 125\,keV), DXA requires 185\,GB of peak RSS against 24\,GB for AnuVikar, a factor of $7.77\times$. By contrast, the Savi-Bhransha graph-analysis step
itself adds only $\approx$307\,MB (median across HCP), confirming that
memory scales with defect count rather than simulation-cell size. These
ratios mean DXA is impractical on standard workstation hardware at the
largest cascade sizes, whereas the Savi-Bhransha workflow remains feasible.

\section{Conclusions}
\label{sec:conclusions}

Savi-Bhransha provides loop-level characterisation directly from defect-core geometry and extends naturally to both interstitial and vacancy populations. Across the three datasets studied here---single-PKA BCC~W, single-PKA HCP~Zr, and successive-cascade FCC~FeNiCr---it returns habit plane, Burgers vector, dumbbell orientation, mixed-defect morphology over a wide size range, and segment-wise edge/screw character within a single occupancy-based pipeline, and reproduces or quantitatively brackets several experimentally accessible observables. The native separation of boundary and bulk defect populations, in particular, recovers the TGS-measured boundary-defect concentration in tungsten and supports a boundary-driven interpretation of the thermal-diffusivity signal.

Methodologically, Savi-Bhransha realises the inverse of the classical forward dislocation construction~\cite{mura1987,barnett1985,cai2006nonsingular}---recovering loop topology from the displacement-field core rather than imposing it on the lattice---and is therefore on a distinct theoretical lineage from interface-mesh Burgers-circuit methods. Against DXA, the method reproduces the physically meaningful total-length trends while providing a substantially more stable loop-count observable when loops are fragmented or embedded in interacting defect environments: closed-loop counts alone do not correlate with Savi-Bhransha across the 64-cascade filtered benchmark, whereas the loops-plus-segments observable restores strong agreement. The qualitative comparison further indicates that the divergences between the two methods are systematic enough to be characterised rather than averaged out, and that count-, density-, and sign-based observables are best derived from the loop-level Savi-Bhransha output. The timing and peak-memory advantages over DXA grow with system size and approach an order of magnitude across the tested cascade boxes, making the workflow practical on standard workstations at scales where DXA becomes infeasible.

The present implementation also exposes the next development priorities clearly: more rigorous fresh-process memory benchmarking for exact cross-tool comparison, explicit treatment of stacking-fault-rich loops, and continued extension of vacancy-loop handling in mixed defect populations. These are natural extensions of the same defect-core graph formalism rather than separate methodological branches. We anticipate that the framework will be applied to a wider range of materials and defect environments in future work, and that the line-primitive view it provides will support more mechanistically informed analysis of cascade microstructures than is accessible from a global mesh alone.

\section*{Acknowledgments}
The author thanks Dr.~Manoj Warrier, Shri Aditya Majalee, and Shrimati Poonam Pahari for providing the molecular-dynamics cascade simulation data used in this work, and acknowledges the computational resources provided by the BARC Computer Centre.


\bibliographystyle{unsrtnat}
\bibliography{refs}

\end{document}